\documentclass[useAMS,usenatbib]{mn2e}
\usepackage{graphicx}
\usepackage{latexsym,epsfig}

\title[Arcturus stream and AF06 stream]{Chemical compositions of stars in two stellar streams from the
Galactic thick disk}
\author[Ramya, Reddy and Lambert]{P. Ramya,$^{1}$\thanks{E-mail:
ramyap@iiap.res.in} Bacham E. Reddy$^{1}$ and David L. Lambert$^{2}$ \\
$^{1}$Indian Institute of Astrophysics, Bengaluru, India-560034\\
$^{2}$University of Texas, McDonald Observatory, Austin, TX, USA}

\begin{document}

%\date{Accepted ..... Received .... ; in original form ....}

\pagerange{\pageref{firstpage}--\pageref{lastpage}} \pubyear{...}

\maketitle

\label{firstpage}

\begin{abstract}
We present abundances for 20 elements for stars in two stellar streams
identified by \citealt{arif06}: 18 stars from the Arcturus stream and 26 from
a new stream, which we call AF06 stream, both from the Galactic thick disk.
Results show both streams are metal-poor and very old (10$-$14 Gyrs) with  
kinematics and abundances overlapping with the properties of local
field thick disk stars. Both streams exhibit a range in metallicity
but with relative elemental abundances that are identical to those of
thick disk stars of the same metallicity. 
These results show that neither stream can
result from dissolution of an open cluster. It is highly unlikely that
either stream represents
 tidal debris from an accreted satellite galaxy. Both streams
most probably owe their origin to dynamical perturbations within the
Galaxy.   
 
\end{abstract}


\section{Introduction}
The phase space of the Galaxy's disk, as observed near the
Sun, contains substructure within the larger structures known as
the thin and the thick disk.
Substructures include
open clusters and stellar streams, also often referred to as
moving groups. Moving groups, as promoted
by Eggen (see e.g., \citealt{eggen98} and references therein),
are considered to be stars having 
a common space motion and a common chemical
composition, and originating from a disssolving open cluster.
While some of the well known
moving groups do share a common composition (e.g., the HR 1614 group -
\citealt{desilva07}) suggesting they come from a tidally-disrupted
open cluster, other groups contain stars having very different compositions,
a result demanding an origin more complex  than
disruption of an open cluster. The term `stream' is now applied to
some entities in Galactic phase space. Examples in the thin disk include
the Hercules stream \citep{bensby07} and
the Hyades stream or supercluster (\citealt{desilva2011}, \citealt{pomp2011}).
For streams in the thin disk, a likely explanation
involves dynamical interactions of disk stars with the central
bar \citep{antoja09} or spiral density waves \citep{minchev2010}. Other possibilities arise for streams belonging to the
thick disk including accretion from external galaxies.

In this paper, we present chemical compositions for subdwarfs belonging to
two streams in the thick disk.
\cite{arif06} undertook a
search for fine structure in the phase space populated by
subdwarfs from the large sample of F and G subdwarfs
considered by \cite{carney94} for which \cite{arif06}
refined data on stellar distances and kinematics. Two
clumps in phase space were noted by \cite{arif06}.
One with $V$ $= -125$ km s$^{-1}$ and $\sqrt{U^2 + 2V^2} =
185$ km s$^{-1}$ is referred to as the Arcturus stream.
An Arcturus moving group had been
previously identified by \cite{eggen71}. The second stream AF06
with a stronger presence in phase space than the Arcturus
stream is at $V = -80$ km s$^{-1}$ and $\sqrt{U^2 + 2V^2} =
130$ km s$^{-1}$.

Here, we report chemical compositions of stars from
these two streams. We show
that stars in both streams span a range in metallicity but
with relative abundances which match
closely the ratios reported for field thick
disk stars. This result serves to constrain greatly explanations
for the origins of the two streams.

\section{Samples Stars and Observations}

Stars selected for observation came from membership lists
of the Arcturus and AF06 streams given by \cite{arif06}:
18 of the 22 stars in former stream and 26 of the 44 stars
in the latter stream were observed successfully with the
Tull coud\'{e} spectrograph \citep{tull95} at the
2.7 meter Harlan J. Smith telescope of the W.J. McDonald
Observatory.

Spectra at a resolving power of 60,000 were obtained with
spectral coverage from about 3800 \AA\ - 10000 \AA\  with echelle
orders incompletely recorded on the CCD beyond about 5800 \AA.
Wavelength calibration was provided by an exposure of a Th-Ar hollow
cathode lamp.
These two-dimensional data were reduced to one-dimensional
relative flux versus wavelength spectra using the Image Reduction and
Analysis Facility \footnote{IRAF is distributed by the National Optical Astronomy
Observatory, which is operated by the Association of Universities for
Research in Astronomy (AURA) under cooperative agreement with the National
Science Foundation.}(IRAF).
In a typical spectrum, the S/N ratio at the centre of an
order over most of the full spectral range was 100 or greater.

\section {Analysis}
\subsection{Stellar Atmospheric Parameters}

Atmospheric parameters $-$ effective temperature (T$_{\rm eff}$), surface
gravity (log $g$) and metallicity ([M/H]) have been derived from both photometric and spectroscopic data.
In the former case, we relied on published catalogues of photometry and
parallaxes, empirical calibrations and the theoretical
stellar evolutionary models. And, in the latter case, our high
resolution spectra were used to derive the atmospheric
parameters including microturbulence ($\xi_{\rm t}$). Below, both the procedures are
described in brief.

\subsubsection{Photometry}

The T$_{\rm eff}$ is derived using (V-K$_{\rm s}$) colour and Str\"{o}mgren photometry
($uvby$) calibrations. The K$_{\rm s}$ magnitude is taken from 2MASS
catalogue\footnote{This publication makes use of data products from the 2MASS, which is
a joint project of the University of Massachusetts and the Infrared Processing and
Analysis Centre/California Institute of Technology, funded by the
National Aeronautics and Space Administration (NASA) and the National
Science Foundation} \citep{cutri03}.
The subscript "s" stands for the  bandpass of the K filter in the 2MASS
survey, i.e., the K$_{\rm s}$ filter is narrower than the Johnson K filter.
The K$_{\rm s}$ magnitudes are converted to
standard "K" magnitudes using
relations given in \cite{ramirez05}. 
The mean difference between the two magnitudes is only
K$_{\rm s}$ $-$ K$_{\rm tcs}$ = $-$0.001 $\pm$ 0.005
and will have no effect when K$_{\rm s}$ is used in place of K
magnitudes in the calibration  between (V-K) and  T$_{\rm eff}$.
The V magnitudes for all the stars
were adopted from \cite{kharch01}.
The (V-K) colour and the empirical relations
provided in \cite{alonso96} are used in deriving T$_{\rm eff}$.
Str\"{o}mgren colours and indices
 ($b-y$, $m_{1}$, $c_{1}$) are 
available for 26 out of 44 stars in the sample \citep{hauck98}.
Values of metallicity and T$_{\rm eff}$
were obtained using empirical calibrations of Str\"{o}mgren colours and
indices given in \cite{schuster89}
and \cite{alonso96}, respectively. Values of metallicity are quite sensitive
to reddening as it makes observed $(b-y)$ more positive and $m_{1}$ values more
negative than their intrinsic colours. 
However, we expect no significant reddening as the stars are nearby (d $<$130 pc from the Sun).
Using the methods given in \cite{schuster89}, reddening values E($b-y$) 
have been estimated and, indeed, all reddening estimates are
 vanishingly small:
E($b-y$) $\leq$ 0.001$\pm$0.006.
Temperatures derived using
(V-K$_{\rm s}$) and Str\"{o}mgren colours are given in Table~1 as
(T$_{\rm eff}$)$_{\rm V-K}$ and (T$_{\rm eff}$)$_{\rm b-y}$,
respectively.
The mean difference
between the two temperatures, (T$_{\rm eff}$)$_{\rm V-K}$ $-$ (T$_{\rm eff}$)$_{\rm b-y}$ = 18 $\pm$ 90 K, excluding the
outliers HIP24030, HIP53070, HIP11952 and G192$-$21, for which the
difference is large 245 $\pm$ 24 K, i.e.,
T$_{\rm eff}$ derived from (V-K$_{\rm s}$) colour are hotter than those from $(b-y)$. The $(b-y)$
temperature of HIP 24030 and (V-K$_{\rm s}$) temperature of G192$-$21 are much closer to the values
obtained using spectroscopy. In the case of the 
other two, we suspect errors in one of the
their colours.

The
log $g$ value is derived from the trignometrical parallax, the
(B-V) colour, and theoretical isochrones \citep{Demarque04}.
Errors in the  parallax and (B-V) colour are taken into account in
estimating the uncertainty in
the log $g$ value.

\subsubsection{Spectroscopy}

A full set of atmospheric parameters (T$_{\rm eff}$, log $g$, $\xi_{\rm t}$, [M/H])
has been derived from spectral line analysis by standard LTE
techniques. In this exercise, the LTE Kurucz grid of ATLAS9 model atmospheres
with the convective overshoot option was adopted
\citep{kurucz1998}. The rationale for
choosing overshoot models for solar type dwarf stars was given in \cite{reddy03,reddy06}.
Since we intend to compare our results with the results from the thin and thick
disk studies of Reddy et al., we followed their analysis techniques.
The LTE line analysis code MOOG \citep{sneden1973} in its 2009 version was used throughout.

The effective temperature was set by the requirement that the Fe
abundance provided by Fe\,{\sc i} lines be independent of the lower
excitation potential (LEP) of the selected lines. While deriving T$_{\rm eff}$,
caution was taken to minimise the effect of microturbulence by choosing, initially, very
weak lines with a sufficient range in LEP.
Later, microturbulence ($\xi_{\rm t}$) was derived
by adding  Fe\,{\sc i} lines of moderately strong equivalent width
 (W$_{\lambda}$ $\leq$ 120 m\AA)
so that the abundance trend becomes sensitive to changes in $\xi_{\rm t}$.
The chosen value of $\xi_{\rm t}$ is that for which the
abundance is independent of equivalent width. The
surface gravity log $g$ was obtained by requiring that, for the given T$_{\rm eff}$
and $\xi_{\rm t}$, Fe\,{\sc i} and Fe\,{\sc ii} lines give the same
Fe abundance. The uncertainties in the derived parameters
have been estimated by inspection of dependencies
for combinations of models of different sets of parameters. In the
case of T$_{\rm eff}$, we varied the best representative T$_{\rm eff}$ 
in steps of 25~K for given log $g$, $\xi_{t}$ and
[M/H]. For steps of 25~K changes, we found no significant
changes in the slope as well as in abundances, however, we see (See Figure~2) 
noticeble changes in abundance trends
by increasing or decreasing 50~K from its mean model T$_{\rm eff}$. Thus, we estimate
$\pm$50 K as an uncertainty in the best fit model atmosphere.
Similarly, we found model uncertaities in log $g$ and $\xi_{t}$. In Figure~2,
estimation of uncertainties is illustrated for T$_{\rm eff}$ and $\xi{t}$.
In this way, we found model uncertainties in T$_{\rm eff}$, log $g$, and
$\xi_{t}$: $\pm$ 50~K, $\pm$ 0.20 cm s$^{-2}$, $\pm$ 0.20 km s$^{-1}$, respectively.
These individual uncertainties translate to an effective error
of $\pm$ 0.05 dex in metallicity [Fe/H].

Next, photometric and spectroscopic estimates of model atmosphere parameters are
compared.
Temperature, log $g$ and metallicity comparisons are given in
Figure~1. The mean difference between (T$_{\rm eff}$)$_{\rm V-K}$ and spectroscopic
T$_{\rm eff}$ is just -9 $\pm$ 87~K.
There are a few outliers HIP24030 and HIP94931 (for which (T$_{\rm eff}$)$_{\rm b-y}$ 
values are closer to spectroscopic T$_{\rm eff}$), G10$-$12 and HIP9080 (for which (T$_{\rm eff}$)$_{\rm b-y}$ not available) 
with a difference of -77 $\pm$ 174~K. Excluding the outliers
the mean difference becomes -2 $\pm$ 73~K.
The log $g$ values obtained from photometry are in good agreement
with the spectroscopic values: the difference between the two methods
is only  0.002 $\pm$ 0.18 cm s$^{-2}$
for 23 stars for which photometric gravities could be obtained, 
a difference
within the combined uncertainties.
The mean difference
between the photometric metallicity from Str\"{o}mgren photometry and the
 spectroscopic metallicity
is $-$0.03 $\pm$ 0.11 dex for 25 stars for
which Str\"{o}mgren photometry
is available. Star G192$-$21 is an outliers whose spectroscopic metallicity
is about 0.7 dex metal richer compared to the corresponding photometric value.
Difference is probably due to erroneous Str\"{o}mgren photometry. 
For the two metal-poor stars HIP53070 and HIP11952, too few lines are
available for a spectroscopic determination of parameters and, therefore,
we have adopted
photometric (T$_{\rm eff}$)$_{\rm b-y}$ and log $g$ values.
The microturbulent velocity $\xi_{\rm t}$ calculated using the relation given
in \cite{reddy03} between
$\xi_{\rm t}$, T$_{\rm eff}$, and log $g$ was adopted
for both stars. In the final calculations of abundances we adopt
parameters derived from spectroscopy but
no conclusions would be changed were the
photometric parameters adopted.

\subsection{Abundances}

Elemental abundances have been derived using measured equivalent widths and
synthetic spectra with LTE model atmospheres for the adopted stellar parameters 
taken from Kurucz grid \citep{kurucz1998} and the 2009 version of LTE line analysis code
MOOG. The line list
compiled by \cite{reddy06}
was adopted. Solar abundances derived by \cite{reddy03}
have been used as rereference values, except Eu abundance. The solar 
Eu abundance (log $\varepsilon$(Eu) = 0.55) has been derived, 
for this study, using the Atlas solar spectrum \citep{hinkle00}
and two Eu II lines (6645.13 \AA, 4129.72 \AA).
For transitions with significant hyperfine structure (HFS)
data were taken from Kurucz HFS data base \citep{kurucz1998}. The lines of Mn,
V, Cu, and Eu with HFS were analyzed by computing synthetic
spectra.
In the case of vanadium, the line at 6216.36 \AA,
one of the six lines used previously, is now judged to
be blended in the cooler stars and is omitted from the analysis of such stars.
Elemental abundances are summarized in Tables 2 and 3.

Uncertainties in the derived abundances arise primarily
from uncertainties in the derived atmospheric parameters,
the equivalent widths (W$_{\lambda}$) and oscillator strengths (log $gf$).
The latter two quantities affect primarily the line-to-line scatter.
Of course, the line-to-line scatter is to some extent
also influenced by uncertain model parameters. Uncertainty in the log $gf$
values is minimized as our analysis is a differential one with respect to
solar spectrum and photospheric abundances.
Error in the measured W$_{\lambda}$ is estimated following
the recipe given in \cite{cayrel1988}. For the quality of our data, uncertainty
in the measured W$_{\lambda}$ varies from 1m{\AA} to about
2.5m{\AA}. We have taken on average 2m{\AA} as the uncertainty in the
measured W$_{\lambda}$.
In Table~4, we have provided for one representative star (HIP40613), estimated
uncertainties for each element in the form of $\Delta$[X/Fe] due to $\delta$T$_{\rm eff}$,
$\delta$log $g$, $\delta${$\xi_{\rm t}$, $\delta$[M/H], and $\delta$W$_{\lambda}$.
Assuming all five sources of error discussed above are independent,
the combined error is obtained by adding them
 in quadrature. This we call our measured uncertainty or ${\sigma_{\rm model}}$.

Uncertainties in abundances can also be gauged by comparing
results for stars that are common in this and other studies.
For seven stars that are common with thick disk sample study by \cite{reddy06},
differences in the derived abundance ratios [X/Fe] between the studies are
given in Table~5. The differences are quite small and less
than measured uncertaintity (${\sigma_{\rm model}}$). The good agreement between
the two studies implies that the results of the two streams in this study
can be compared directly with the results of thick disk sample study from
\cite{reddy06}. This is akin to a differential analysis with respect to thick disk.

\subsection {Kinematics}

Both streams were identified by \cite{arif06} as over-densities of stars
in phase space. To represent stars in phase
space one requires information such as coordinates, radial velocities, distance (parallaxes),
and proper motions.
In the calculation of Galactic
motions \cite{arif06} used Hipparcos data \citep{perry97} and radial
velocities from \cite{carney94}. Recently, Hipparcos observations were re-reduced by
\cite{van07}. 
Consequently, we have
rederived the Galactic velocities (U, V, W) for stream members
using the revised Hipparcos parallaxes and proper motions and including
the radial velocities derived from our McDonald spectra.
Radial velocities obtained in this study are in very good agreement with the velocities
given in \cite{arif06}. The mean difference between the two studies is 0.15 $\pm$ 0.50 km s$^{-1}$.

The U, V and W velocities with respect to the Sun  
were calculated using the method given in \cite{john87}. A right-handed 
coordinate system is used throughout 
where U is positive towards the 
Galactic center, V is positive in the direction of Galactic rotation and W
is positive towards the North Galactic Pole (NGP). 
Velocities (U, V, W) are in good agreement with values given by \cite{arif06};
differences between the two 
studies are mainly due to the updated parallaxes.
None of the stars identified as members by \cite{arif06} lost their membership
in either stream.
Derived radial velocities (R$_{\rm v}$) and 
Galactic velocities (U$_{\rm LSR}$, V$_{\rm LSR}$, W$_{\rm LSR}$) relative to Local Standard of Rest
(LSR) are given in Table~6. In conversion to the LSR frame, the solar
motion of (U$_{\odot}$, V$_{\odot}$, W$_{\odot}$) = ($+$10.0, $+$5.3, $+$7.2) km s$^{-1}$) is used \citep{dehnen98}.  

The mean motion (U$_{\rm LSR}$, V$_{\rm LSR}$, W$_{\rm LSR}$) of the
Arcturus stream based on the 18 member stars observed by us
is ($-$6.48 $\pm$ 49.29, $-$124.79 $\pm$ 8.92, $-$11.5 $\pm$ 49.59)
and for AF06 stream the mean motion is
($-$41.55 $\pm$ 47.45, $-$87.35 $\pm$ 7.83, 3.82 $\pm$ 54.34).
These values are in good agreement with the streams'
central values given in \cite{arif06}.

To compute orbital parameters, the (U, V, W) of each star is integrated 
over the Galactic potential    
provided by D. Lin (private communication).
Orbital parameters: mean of apogalactic and perigalactic distance (R$_{\rm m}$), 
eccentricity (e), maximum distance star away from the Galactic 
plane (Z$_{\rm max}$) have been derived and given in Table~6. 
A distance of 8.5 kpc beween the Sun and the Galactic center is used in the
calculation. 

The (U,V,W) of both the streams, as \cite{arif06} appreciated,
suggest that the streams are part of the thick disk. 
Probabilities that a particular star belongs to the halo, the thick or the thin disk
are calculated using the definitions and recipes given in
\cite{reddy06}. The percentage probability (P) that a star
belongs to the thick disk is given in Table~6.
All the Arcturus stream members in our sample are thick disk stars with a probability
P$_{\rm Thick}$ $\geq$ 80 $\%$, while 15 out of
the 26 AF06 stream members have a probability that would qualify them 
as thick disk members  with the remaining 11 stars having probabilities placing them in either
the thin or thick disks.

In Figure~3, sample stars of the two streams along with the representative
members of the thick disk, the thin disk and the halo are shown in the space of 
angular momentum per unit mass components J$_{z}$, and J$_{\bot}$= $\sqrt{J_{\rm x}^2+J_{\rm y}^2} $ 
the 
azimuthal and perpendicular components, respectively. 
In angular momentum space, stars  are
clustered in a small region compared to their distribution
 in the velocity space (see \cite{helmi00}).
Values are calculated in the right-handed coordinate system and 
the LSR velocity is assumed to be 220 km s$^{-1}$. 
The value J$_z$ = 0 implies no rotational velocity
in the direction of the Galactic rotation and increasing values
mean higher rotational velocities. 
The J$_{\bot}$ represents extent of tilt of star's
orbit with respect to the Galactic plane. Obviously, for the thin disk it is quite small.
 Arcturus stream and AF06 stream have a mean J$_{\rm z}$ of 
$-$811 $\pm$ 77 kpc km s$^{-1}$
and $-$1130 $\pm$ 63 kpc km, s$^{-1}$, respectively. 

The presence of the Arcturus stream
was detected by \cite{navarro04} in  
catalogues of metal-poor stars by \cite{beers00} and
\cite{gratton03}.
They found the Arcturus stream lagging the LSR by 120 km s$^{-1}$ but with a large
dispersion of $\sigma_{\rm V}$ $\sim$ 50 km/s.
Their detected structure has angular momentum (J$_{z}$)
in the range of (700 $-$ 1100) km $s^{-1}$ kpc. 
It appears to us that
this structure detected by \cite{navarro04} is the combined structure of
the Arcturus and the AF06 stream but this structure was separated into two different streams
by \cite{arif06} using the wavelet transform technique.
By combining our samples of both streams,
we get mean values V$_{\rm LSR}$ and J$_{\rm z}$ ($-$102.67$\pm$20.34, $-$1000$\pm$173)
very similar to what \cite{navarro04} found for the
structure that they thought to be the single structure labelled the Arcturus stream.
The Arcturus stream has been identified in recent
studies \citep{klement08} as a significant over density in phase space. 
\cite{klement08}'s study is based on more than 7000 stars within a distance of 500~pc taken from RAVE survey.
They have identified stars with high orbital eccentricities having over-densities at about V= $-$120 km s$^{-1}$ and
the second one at V=$-$95 km s$^{-1}$ which coincide with phase space coordinates
of the Arcturus and AF06 streams as given \cite{arif06}. 
\cite{williams09} examined a data set of 16,000 giants 
from the RAVE survey and showed pronounced over-density of 
stars at velocities overlapping with the
Arcturus stream in the solar circle. 

\subsection{Ages}

Determination of stellar ages from location of a star in a
colour-magnitude diagram is an uncertain procedure because the
majority of the stars in both streams fall very close to the
zero age main sequence. For the few stars that appear relatively evolved off the
main sequence ages were estimated.  
Mean ages with upper and lower limits are given in Table~6. 
Errors in B-V and parallaxes are
used to estimate the limits. 
In Figure~4, the stars are shown in the HR diagram of M$_{\rm V}$ versus (B-V) colour
along with isochrones of ages 10 Gyrs, 12 Gyrs and 16 Gyrs 
for the metallicities [Fe/H] =$-$0.70 dex 
and [Fe/H] = $-$0.51 dex, for Arcturus stream and AF06 stream respectively.  
Ages of the stream members range from 10 to 14 Gyrs. 
The stellar ages, therefore,
are very similar to the ages of thick disk
field stars which are all older than about 10$-$11 Gyr \citep{reddy06}.
The derived ages are in good agreement
with previous studies of Arcturus group (cf. \citealt{williams09}, 
\citealt{helmi06}, \citealt{navarro04}) based on different selections of
stream members.

\section {Results and Discussion}

\subsection {Chemical signatures}

Since their discovery (cf. \citealt{eggen57}), various theories have been
put forward to explain the substructures in phase space  
known variously as moving groups or stellar streams. 
The most 
prominent theories identify a stellar stream as either
  a) a dissolved open cluster,
b) debris from an accreted satellite galaxy, or  c) the result of
dynamical perturbations within the Galaxy. 

In establishing the correct explanation for a particular stream, the importance
of the chemical signatures or chemical tagging of stream members 
has been recognized (\citealt{freeman02}, \citealt{bensby07}, 
\citealt{williams09}, \citealt{desilva07}) 
not only to pinpoint their origin but also to understand the formation and
evolution of the disk.

Quantitative abundances of twenty elements
has been extracted here for members of the two
streams. 
There are three
main groups of elements: iron peak (V, Cr, Mn, Fe, Ni, Co), 
$\alpha$-elements (O,Mg, Si, Ca, Ti),
and heavy elements such as $s$-process (Ba, Nd) and $r$-process (Eu) elements. 
For the present (thin) disk, iron peak elements 
mostly come from Type Ia supernovae explosions (SNIa) and
 the $\alpha$-elements are primarily produced in Type II supernovae (SNII).
The $s$-process elements are known to be produced mainly in evolved stars of 
low and intermediate
mass (1 $-$ 8M$_{\odot}$) with the lighter $s$-process elements also contributed by
SNII. Eu, an $r$-process element, is
most proably produced in SNII.
Therefore, an abundance ratio of products of
SNIa and SNII helps to track the chemical history of a stellar system.

Abundances of elements from O to Eu in the form of [X/Fe] ratios, where X is any element, 
are given in Table~2 and ~3 and 
shown in Figures~5, 6, and 7 as the run of [X/Fe] against [Fe/H]. 
The metallicity distribution of the streams
is shown in Figure~8 for the 
Arcturus and AF06 streams.
Next, we discuss these abundances for  
the Arcturus and AF06 streams in the light of the proposed scenarios for
the origin of stellar streams.

\subsection{Dissolved Stellar Clusters}

Dispersal of an open cluster as the origin of a stellar stream is 
one explanation from the suite of potential explanations that is directly
testable from photometric and spectroscopic determinations of
stellar compositions. Given that open clusters exhibit chemical
homogeneity, if a stream represents a dissolved cluster, one requires
chemical homogeneity also among stream members.

Chemical homogeneity is not a characteristic of these two streams. Results shown
in Figure~8
and Table~2 show that stellar metallicities
span a wide range for both streams. For the Arcturus stream, [Fe/H] runs from
$-$1.40 to $-$0.37. For AF06 stream, the range is from $-1.69$ to $+0.22$ but
the range is $-$1.69 to $-$0.17 for those 15 members with a high probability of
belonging to the thick disk. For a strictly homogenous populations
such as 
open and globular clusters the degree of chemical homogeneity is quite high and which is
about 
0.05 dex (\citealt{desilva06}, \citealt{pancino2010}) for a number of elements. In the case
of [Fe/H] dispersions are found to be in the range of 0.02 $-$ 0.1 and in some extreme cases
dispersions are of the order of 0.2 dex
\citep{paunzen2010}. 
The wide range in metallicity clearly shows that the
systems from which the stream members originated had a relatively long history with
a multiple episodes of star formation. 

 We have also inspected the sample
for an evidence of a sub-group with chemical homegenity. Results shown in Figure~5, 6 and 7
indicate existence of no such a group among Arcturus stream sample,  
however, for the stream AF06, we find 
a hint of clustering of stars at [Fe/H] = $-$0.4. About 8 stars (out of total 26) show
metallicity [Fe/H] = $-$0.4 within the dispersion of 0.04 dex. 
Does it mean that part of the sample stars originated 
from the disrupted cluster? Probably not, this may be a manifestation of 
thick disk metallicity distribution
which peaks at about $-$0.6 dex, and to some extent due to a smaller sample size. 
Thus, neither the Arcturus nor the AF06 stream, as shown
by metallicity distribution of member stars of the two streams (see Figure~8),
is a dissolved open cluster. This conclusion about the
Arcturus stream was reached also by \cite{williams09} who selected
134 stream members by selection criteria different from those used by
\cite{arif06} and applied to stellar catalogues other than that
compiled by \cite{carney94}. 
Their analyses of high-resolution spectra led to a [Fe/H] range similar to
that quoted above (see also \cite{navarro04}).

\subsection{Are the streams chemically identical to thick disk field stars?}

Among thick disk field stars in the solar neighborhood,
there is a strikingly very small dispersion in elemental
abundance ratios, i.e., [X/Fe] at a given [Fe/H]. Indeed, \cite{reddy06}
found the [X/Fe] to be Gaussian-like with a dispersion $\sigma$ of less
than 0.10 dex for the most of the elements except for V, Y, and  Zr for which
$\sigma$ is slightly more than 0.1 dex. Furthermore, such dispersions
were uncorrected for measurement uncertainties so that the
intrinsic or `cosmic' dispersion must be very small.

In sharp contrast to the very similar [X/Fe] ratios
at a given [Fe/H] for local field stars, element ratios
reflecting different contributions from major processes of stellar
nucleosynthesis do vary from stellar system to stellar system. For example,
ratios in the Galactic bulge are not uniformly identical at a given [Fe/H]
to those among local thick or thin disk field stars. Similarly, ratios of
$\alpha$-elements ([$\alpha$/Fe]) at a given [Fe/H] among stars of
dwarf spheroidal galaxies differ appreciably from those of local stars 
and from galaxy to galaxy (\citealt{venn04}, \citealt{tolstoy09}, \citealt{kirby11}).
This has been illustrtated in Figure~9 where ratio of [$\alpha$-process/Fe] (mean of $\alpha$-process
elements Mg, Si, Ca and Ti) is compared with that of thick disk \citep{reddy06} 
and a number of dwarf spheroidal galaxies
for which data was taken from
(\citealt{venn04}, \citealt{monaco05}).

Therefore, the chemical signatures or tags in the form of [X/Fe] for
those key elements from the major processes of nucleosynthesis may test
some proposed origins for the streams.
To address the question `Are these two streams chemically identical
to thick disk field stars?', we show plots of [X/Fe] versus [Fe/H] 
in Figures 5 to 7 with field stars
from \cite{reddy06} shown as by grey symbols, and the Arcturus and
AF06 stream members shown as black filled circles. To quantify
possible systematic offsets between the field thick disk  and stream members,
we compute mean
values and dispersions for the Arcturus and AF06 streams over the [Fe/H] interval
$-$0.30 to $-$1.0.
Dispersions about the trends is a combination
of both the cosmic scatter as well as errors associated with the model parameters.
The values ($\sigma_{\rm AF06}$, $\sigma_{\rm Arcturus}$)
are computed as the standard deviation of the residuals from
straight line fits to abundance trends of 
[X/Fe] against [Fe/H]. 
Similarly, $\sigma_{\rm Thickdisk}$ for the thick disk abundance trends
of \cite{reddy06} are computed over the same interval. 
All results are provided in
Table~7. In Figure~10, we made comparison of $\sigma_{\rm Thickdisk}$ with those of Arcturus and AF06 streams.
 Dispersion values about the abundance trends of Arcturus stream 
show very good agreement with that
of the thick disk within about 0.02. However, for AF06 stream, in most cases, 
dispersions are lower by 0.01 $-$ 0.03
compared to thick disk dispersions. With the current limited sample it would be far fetching to attribute
this to the different chemical evolution for the AF06 stream and hence external origin to it. 
Mean abundances of Arcturus and AF06 streams are quite similar and the differences are
within 0.05 dex except three elements (O, Cu, Eu) for which difference is 0.06 - 0.07 dex.
Dispersions about the trends for all the elements are comparable to the estimated 
scatter ($\sigma_{\rm model}$) due to uncertainties in model 
parameters. Diferences between dispersions and $\sigma_{\rm model}$} are within 0.05
dex. 
Thus, we conclude that the streams are chemically identical to within
high precision with the
field thick disk stars.
Given that external galaxies of the Local Group have different element
ratios across the [Fe/H] range sampled by these two streams, 
especially for $\alpha$-elements (see, for example, \cite{tolstoy09}, Figure~11), 
it seems most unlikely that either stream represents debris from
an accreted satellite galaxy.

\section{Conclusions}

The outstanding results of our abundance analyses are that the subdwarfs 
comprising the Arcturus stream
and AF06 stream
identified as over-densities in phase space by \cite{arif06} have (i)
a considerable spread in metallicity and (ii) relative abundance [X/Fe]
identical to those of the Galactic thick disk.
The metallicity spread excludes the hypothesis that either stream
represents a dissolved open cluster. The high-degree of similarity
between [X/Fe] at a given [Fe/H] for these streams and the field stars of the
thick disk greatly strains a proposal that these streams represent the tidal
debris of an accreted satellite galaxy. By exclusion, the likely origin
of these streams is that they are the product of dynamical interactions
within the Galaxy.

However, chemical identiy with the thick disk and the non-similarity
of their chemistry with the satellite galaxies within local group alone may not suffice
to rule out the possibility of these streams being a result of disrupted
satellites. It is still a matter of debate how perturbation can create streams with
such a very high velocity drag and exhibit very tight abundance trends. It would be
important to know from the models the extent of regions that get affected due to 
bar/spiral perturbations and their effect on the abundance trends of clumped stars in phase space.
Astrometry from GAIA will provide unprecented sample size as well as accuracy to map the
Galaxy precisely which would decipher
the Galaxy formation and evolution.

\section*{Acknowledgments}
David Lambert thanks the Robert A. Welch Foundation for support through
grant F-634. We thank Dr. Gajendra Pandey for useful comments. We also
thank the referee for his/her remarks.

\begin{table*}
 \centering
 \begin{minipage}{140mm}
  \caption{The atmospheric parameters - Photometry \& Spectroscopy}
  \begin{tabular}{@{}lccccccccc@{}}
  \hline

\multicolumn{5}{c|}{Photometry} & \multicolumn{5}{c}{Spectroscopy} \\ 
Star     &    (T$_{\rm eff}$)$_{\rm V-K}$ &   (T$_{\rm eff}$)$_{\rm b-y}$  & 
log $g$ $\pm$ error   &  ([M/H])$_{\rm b-y}$  &
 T$_{\rm eff}$   &  log $g$  & 
([Fe/H])$_{\rm model}$   &   $\xi_{\rm t}$  &  N\\

    &   K   &   K  &  cm s$^{-1}$  &  dex  & K  &  cm s$^{-1}$  &  dex  &   km s$^{-1}$  &  \\ 
\hline
            &        &       &                      &               &                    &             &         &          \\
\multicolumn{10}{c}{Arcturus stream}\\
            &        &       &                      &               &                    &             &         &          \\
 HIP105888  &   5798 & 5665 &  4.05 $\pm$ 0.05      &    $-$0.81    &   5790   &   4.30  &     $-$0.55  &   1.08  &   (42,7) \\
 HIP36710   &   5301 & ...  &  ...                  &     ...       &   5340   &   4.63  &     $-$0.45  &   0.48  &   (46,6)\\
 HIP77637   &   5478 & 5550 &  4.31 $\pm$ 0.08      &    $-$0.94    &   5580   &   3.73  &     $-$0.85  &   0.90  &   (30,7) \\
 G103$-$53  &   5435 & 5340 &  4.16 $\pm$ 0.09      &    $-$0.66    &   5290   &   4.40  &     $-$0.65  &   0.52  &   (42,5) \\
 G72$-$12   &   5041 & 5094 &  ...                  &    $-$0.27    &   5060   &   4.64  &     $-$0.40  &   0.67  &   (43,5) \\
 G4$-$2     &   5258 & 5238 &  ...                  &    $-$0.61    &   5160   &   4.64  &     $-$0.70  &   0.48  &   (37,4) \\
 HIP53070   &   5962 & 5719 &  4.23 $\pm$ 0.04      &    $-$1.30    &   ...    &   ...   &     $-$1.40  &   1.36  &   (4,3)  \\
 G204$-$30  &   5610 & ...  &  ...                  &     ...       &   5550   &   4.42  &     $-$0.80  &   0.69  &   (27,5) \\
 G139$-$49  &   5331 & ...  &  ...                  &     ...       &   5380   &   4.12  &     $-$0.75  &   0.47  &   (30,5) \\
 G241$-$7   &   5446 & ...  &  ...                  &     ...       &   5320   &   4.00  &     $-$0.95  &   0.91  &   (25,4) \\
 HIP40613   &   5723 & 5670 &  4.16 $\pm$ 0.03      &    $-$0.64    &   5670   &   4.02  &     $-$0.55  &   0.90  &   (45,6) \\
 G42$-$34   &   4858 & ...  &  ...                  &     ...       &   4920   &   4.22  &     $-$0.60  &   0.44  &   (47,4) \\
 HIP36491   &   5741 & 5681 &  4.41 $\pm$ 0.05      &    $-$0.96    &   5760   &   4.20  &     $-$0.85  &   1.10  &   (25,6) \\
 HIP94931   &   4964 & 5118 &  4.56 $\pm$ 0.01      &    $-$0.35    &   5120   &   4.58  &     $-$0 40  &   0.55  &   (48,6) \\
 HIP74033   &   5647 & 5574 &  4.02 $\pm$ 0.04      &    $-$0.92    &   5690   &   4.04  &     $-$0.70  &   1.06  &   (42,7) \\
 G5$-$1     &   5562 & ...  &  4.37 $\pm$ 0.08      &     ...       &   5470   &   4.25  &     $-$1.05  &   0.45  &   (24,5) \\
 G102$-$44  &   5253 & ...  &  ...                  &     ...       &   5260   &   4.43  &     $-$0.63  &   0.44  &   (45,5) \\
 HIP58253   &   5359 & 5351 &  ...                  &    $-$0.37    &   5280   &   4.38  &     $-$0.35  &   0.52  &   (41,4) \\
            &        &      &                       &               &          &         &              &         &          \\
\multicolumn{10}{c}{AF06 stream}\\
            &        &      &                       &               &          &         &              &         &          \\
 G67$-$40   &   5473 & 5326 &  ...                  &    $-$0.36    &   5370   &   4.42  &     $-$0.35  &   0.94  &   (48,6) \\
 HIP9080    &   5078 & ...  &  ...                  &     ...       &   5250   &   4.45  &     $-$0.25  &   0.53  &   (51,7) \\
 G66$-$51   &   5420 & ...  &  ...                  &     ...       &   5320   &   4.58  &     $-$0.80  &   0.79  &   (40,5) \\
 G106$-$8   &   5799 & ...  &  4.13 $\pm$ 0.10      &     ...       &   5780   &   4.23  &     $-$0.40  &   1.00  &   (38,8) \\
 HIP10652   &   5607 & 5499 &  4.35 $\pm$ 0.03      &    $-$0.74    &   5580   &   4.42  &     $-$0.60  &   1.02  &   (38,5) \\
 HIP22020   &   5610 & ...  &  4.12 $\pm$ 0.08      &     ...       &   5690   &   4.15  &     $-$0.20  &   0.95  &   (51,8) \\
 HIP26452   &   5837 & ...  &  ...                  &     ...       &   5830   &   4.14  &     $-$0.68  &   0.70  &   (28,6) \\
 HIP31740   &   5293 & 5436 &  4.31 $\pm$ 0.12      &    $-$0.32    &   5430   &   4.45  &     $-$0.35  &   0.62  &   (44,7) \\
 HIP102923  &   4850 & 4933 &  4.61 $\pm$ 0.04      &    $-$0.19    &   4950   &   4.50  &     $-$0.25  &   0.69  &   (49,6) \\
 G146$-$76  &   5170 & ...  &  ...                  &     ...       &   5090   &   2.67  &     $-$1.60  &   0.82  &   (16,5) \\ 
 HIP34642   &   5775 & ...  &  4.07 $\pm$ 0.05      &     ...       &   5800   &   4.07  &     $-$0.40  &   1.00  &   (41,8) \\
 G10$-$12   &   4957 & ...  &  ...                  &     ...       &   5120   &   3.98  &     $-$0.45  &   0.84  &   (50,7) \\
 HIP17147   &   5741 & 5722 &  4.22 $\pm$ 0.03      &    $-$0.83    &   5700   &   4.23  &     $-$0.85  &   0.71  &   (37,6) \\
 HIP24030   &   5915 & 5697 &  4.16 $\pm$ 0.11      &    $-$1.06    &   5730   &   4.20  &     $-$1.05  &   1.16  &   (15,6) \\
 HIP11952   &   6029 & 5785 &  4.24 $\pm$ 0.10      &    $-$1.57    &   ...    &   ...   &     ...      &   1.38  &    (4,4) \\
 HIP29814   &   5160 & 5217 &  4.48 $\pm$ 0.01      &    $-$0.37    &   5230   &   4.48  &     $-$0.40  &   0.75  &   (51,6) \\
 G197$-$45  &   5176 & ...  &  ...                  &     ...       &   5250   &   4.02  &     $-$0.60  &   0.85  &   (38,6) \\
 HIP104913  &   5355 & ...  &  4.40 $\pm$ 0.02      &     ...       &   5380   &   4.48  &     $-$0.03  &   0.80  &   (51,8) \\
 G192$-$21  &   5790 & 5513 &  4.32 $\pm$ 0.10      &    $-$1.23    &   5820   &   4.20  &     $-$0.50  &   0.88  &   (36,6) \\
 G69$-$21   &   5562 & 5500 &  ...                  &    $-$0.30    &   5620   &   4.29  &     $-$0.20  &   1.07  &   (48,7) \\
 G68$-$10   &   5680 & 5589 &  4.36 $\pm$ 0.06      &    $-$0.57    &   5570   &   4.22  &     $-$0.50  &   0.73  &   (47,6) \\
 G30$-$46   &   5076 & ...  &  ...                  &     ...       &   5150   &   4.65  &     $+$0.15  &   0.68  &   (48,6) \\
 HIP16169   &   5638 & 5575 &  4.34 $\pm$ 0.02      &    $-$0.56    &   5690   &   4.56  &     $-$0.48  &   1.10  &   (38,7) \\
 G6$-$16    &   5786 & 5655 &  4.16 $\pm$ 0.10      &    $-$0.19    &   5800   &   4.30  &     $-$0.05  &   1.10  &   (48,8) \\
 G78$-$41   &   5411 & 5494 &  4.37 $\pm$ 0.12      &    $-$0.38    &   5480   &   4.40  &     $-$0.40  &   0.74  &   (50,6) \\
 G25$-$5    &   5487 & 5508 &  ...                  &    $-$0.36    &   5560   &   4.50  &     $-$0.35  &   0.93  &   (50,7) \\

\hline
\end{tabular}
\end{minipage}
\end{table*}

\begin{table*}
 \centering
 \begin{minipage}{140mm}
  \caption{The Abundance ratios ([X/Fe]) of the programme stars}
  \begin{tabular}{@{}lcrrrrrrrrr@{}}
  \hline
Star  &  [FeI/H]  &   [O/Fe]    &   [Na/Fe]   &  [Mg/Fe]  
 &   [Al/Fe]  &  [Si/Fe]   &  [Ca/Fe]   &  [Sc/Fe]  &  [Ti/Fe]  &  [V/Fe]   \\
\hline
          &            &          &             &           &          &           &          &           &          &             \\ 
\multicolumn{11}{c}{Arcturus stream} \\
          &            &          &             &           &          &           &          &           &          &             \\ 
HIP105888 &  $-$0.56    &   0.58  &    0.10    &    0.25  &   0.20  &   0.16   &  0.12   &    0.17  &   0.24  &    0.08 \\
HIP36710  &  $-$0.42    &   0.32  &    0.03    &    0.02  &   0.16  &   0.10   &  0.10   &    0.16  &   0.23  &    0.18 \\
HIP77637  &  $-$0.82    &   0.69  &   $-$0.04  &    0.38  &   0.17  &   0.22   &  0.17   & $-$0.09  &   0.15  & $-$0.06 \\
G103$-$53 &  $-$0.65    &   0.61  &    0.00    &    0.03  &   0.18  &   0.19   &  0.17   &    0.11  &   0.17  &    0.07 \\
G72$-$12  &  $-$0.37    &   0.29  &    0.05    & $-$0.01  &   0.12  &   0.14   &  0.01   &    0.16  &   0.13  &    0.21 \\
G4$-$2    &  $-$0.70    &   0.38  &   $-$0.07  &    0.10  &   0.13  &   0.17   &  0.06   &    0.21  &   0.16  &    0.06 \\
HIP53070  &  $-$1.40    &   0.84  &    ...     &     ...  &    ...  &    ...    &  0.27  &     ...  &   ...   &     ... \\
G204$-$30 &  $-$0.81    &   0.60  &    0.11    &    0.30  &   0.19  &   0.22   &  0.19   &    0.11  &   0.24  &    0.07 \\
G139$-$49 &  $-$0.75    &   0.69  &   $-$0.18  &    0.37  &   0.22  &   0.09   &  0.12   &    0.03  &   0.12  & $-$0.18 \\
G241$-$7  &  $-$0.94    &   0.92  &    0.22    &    0.51  &   0.20  &   0.18   &  0.30   &    0.00  &   0.28  & $-$0.01 \\
HIP40613  &  $-$0.54    &   0.61  &    0.09    &    0.37  &   0.27  &   0.16   &  0.17   &    0.03  &   0.23  &    0.01 \\
G42$-$34  &  $-$0.60    &   0.26  &    0.03    &    0.09  &   0.16  & $-$0.05  &  0.24   &    0.03  &   0.29  &    0.24 \\
HIP36491  &  $-$0.86    &   0.56  &    0.11    &    0.43  &   0.27  &   0.19   &  0.19   &    0.12  &   0.19  &    ---   \\
HIP94931  &  $-$0.42    &   0.39  &   $-$0.01  &    0.12  &   0.14  &   0.11   &  0.10   &    0.18  &   0.21  &    0.23 \\
HIP74033  &  $-$0.70    &   0.60  &    0.09    &    0.35  &   0.20  &   0.16   &  0.13   &    0.11  &   0.19  & $-$0.03 \\
G5$-$1    &  $-$1.04    &   0.62  &   $-$0.05  &    0.31  &   0.30  &   0.00   &  0.11   & $-$0.03  &   0.19  & $-$0.20 \\
G102$-$44 &  $-$0.61    &   0.55  &   $-$0.03  &    0.03  &   0.17  &   0.08   &  0.16   & $-$0.06  &   0.25  &    0.03 \\
HIP58253  &  $-$0.38    &   0.61  &   $-$0.08  &    0.09  &   0.16  &   0.19   &  0.12   & $-$0.05  &   0.15  &    0.18 \\
          &            &          &             &           &          &           &          &           &          &          \\
\multicolumn{11}{c}{AF06 stream}\\
          &            &          &             &           &          &           &          &           &          &          \\
G67$-$40  &  $-$0.38  &   0.37  &    0.08  &    0.21  &   0.22   &   0.18   &  0.16    &    0.08  &   0.21  &    0.22 \\
HIP9080   &  $-$0.17  &   0.24  &  $-$0.03 &    0.08  &   0.04   &   0.03   &  0.01    & $-$0.00  &   0.13  &    0.14 \\
G66$-$51  &  $-$0.80  &   0.46  &  $-$0.01 &    0.27  &   0.21   &   0.14   &  0.21    &    0.03  &   0.29  &    0.18 \\
G106$-$8  &  $-$0.41  &   0.42  &    0.08  &    0.28  &   0.16   &   0.15   &  0.13    &    0.09  &   0.21  &    0.11 \\
HIP10652  &  $-$0.59  &   0.55  &    0.09  &    0.29  &   0.30   &   0.22   &  0.22    &    0.04  &   0.25  &    0.06 \\
HIP22020  &  $-$0.22  &   0.36  &    0.01  &    0.21  &   0.21   &   0.14   &  0.10    &    0.09  &   0.22  &    0.01 \\
HIP26452  &  $-$0.70  &   0.46  &    0.15  &    0.23  &   0.20   &   0.11   &  0.11    & $-$0.05  &   0.16  &    0.02 \\
HIP31740  &  $-$0.36  &   0.47  &    0.10  &    0.13  &   0.20   &   0.16   &  0.15    &    0.19  &   0.27  &    0.14 \\
HIC102923 &  $-$0.29  &   0.32  &    0.14  &    0.11  &   0.06   &   0.11   &  0.12    &    0.16  &   0.29  &    0.47 \\
G146$-$76 &  $-$1.59  &   0.86  &     ...  &    0.28  &    ...   &   0.31   &  0.12    & $-$0.20  & $-$0.01 &     ...  \\
HIP34642  &  $-$0.42  &   0.38  &    0.03  &    0.25  &   0.20   &   0.11   &  0.10    &    0.10  &   0.17  &    0.09 \\
G10$-$12  &  $-$0.43  &   0.55  &    0.07  &    0.25  &   0.22   &   0.19   &  0.18    &    0.20  &   0.30  &    0.18 \\
HIP17147  &  $-$0.86  &   0.75  &    0.08  &    0.42  &   0.28   &   0.22   &  0.22    &    0.08  &   0.23  & $-$0.03 \\
HIP24030  &  $-$1.07  &   0.73  &    0.06  &    0.37  &   0.27   &   0.24   &  0.24    & $-$0.07  &   0.26  &     ...  \\
HIP11952  &  $-$1.69  &   1.25  &     ...  &     ...  &    ...   &    ...   &  0.33    &     ...  &    ...  &     ...  \\  
HIP29814  &  $-$0.40  &   0.36  &    0.02  &    0.04  &   0.16   &   0.11   &  0.12    &    0.08  &   0.20  &    0.16 \\
G197$-$45 &  $-$0.58  &   0.48  &    0.11  &    0.11  &   0.24   &   0.16   &  0.24    & $-$0.10  &   0.24  &    0.10 \\
HIC104913 &  $-$0.02  &   0.16  &  $-$0.05 &    0.09  & $-$0.02  &   0.00   & $-$0.05  &    0.08  &   0.02  &    0.04 \\
G192$-$21 &  $-$0.50  &   0.49  &    0.11  &    0.30  &   0.19   &   0.07   &  0.15    &    0.06  &   0.26  &    0.10 \\
G69$-$21  &  $-$0.22  &   0.27  &    0.01  &    0.12  &   0.11   &   0.04   &  0.04    & $-$0.00  &   0.08  &    0.04 \\
G68$-$10  &  $-$0.50  &   0.56  &    0.08  &    0.14  &   0.21   &   0.14   &  0.19    &    0.01  &   0.25  &    0.03 \\
G30$-$46  &  $+$0.22  & $-$0.03 &  $-$0.09 & $-$0.10  & $-$0.09  & $-$0.05  & $-$0.13  &    0.01  &   0.11  &    0.22 \\
HIP16169  &  $-$0.44  &   0.49  &    0.05  &    0.31  &   0.21   &   0.17   &  0.11    &    0.16  &   0.20  &    0.09 \\
G6$-$16   &  $-$0.02  &   0.09  &  $-$0.08 &    0.07  &   0.02   & $-$0.03  & $-$0.03  &    0.01  &   0.03  & $-$0.03 \\
G78$-$41  &  $-$0.40  &   0.55  &    0.04  &    0.09  &   0.20   &   0.16   &  0.11    &    0.10  &   0.25  &    0.11 \\
G25$-$5   &  $-$0.32  &   0.35  &    0.12  &    0.03  &   0.18   &   0.14   &  0.11    &    0.11  &   0.23  &    0.16 \\
\hline
\end{tabular}
\end{minipage}
\end{table*}

\begin{table*}
 \centering
 \begin{minipage}{180mm}
  \caption{The Abundance ratios ([X/Fe]) of the programme stars}
  \begin{tabular}{@{}lrrrrrrrrrrrr@{}}
  \hline
Star   &   [CrI/Fe]  &  [CrII/Fe]  &   [Mn/Fe]   &   [Co/Fe]   &   [Ni/Fe]  
 &  [Cu/Fe]   &    [Zn/Fe]   &  [Y/Fe]   &  [Ba/Fe]  &   [Ce/Fe]  &  [Nd/Fe]
 &  [Eu/Fe] \\
\hline
            &            &           &             &         &             &          &          &           &             &          &        &        \\ 
\multicolumn{13}{c}{Arcturus stream}\\
            &            &           &             &         &             &          &          &           &             &          &        &        \\ 
HIP105888   &  $-$0.10   &    0.02   &   $-$0.34  &   0.13 &   $-$0.01  & $-$0.05 &   0.30  &    0.02  &   $-$0.17  & $-$0.02 &  0.16 &  0.36 \\
HIP36710    &  $-$0.17   & $-$0.04   &   $-$0.28  &   0.09 &   $-$0.04  &   0.07  &   0.21  & $-$0.08  &   $-$0.22  &   0.15  &  ...   &  0.41 \\
HIP77637    &  $-$0.17   & $-$0.06   &   $-$0.37  & $-$0.01 &  $-$0.09  & $-$0.20 &   0.26  &    0.09  &    0.01    & $-$0.11 &  0.18 &  0.21 \\
G103$-$53   &  $-$0.14   & $-$0.10   &   $-$0.35  &   0.16 &   $-$0.05  &   0.13  &   0.26  & $-$0.11  &   $-$0.19  &   0.14  &  0.26 &  0.10 \\
G72$-$12    &  $-$0.11   & $-$0.04   &   $-$0.24  &   0.10 &   $-$0.06  &   0.09  &   0.17  & $-$0.01  &   $-$0.19  &   0.10  &  0.38 &  0.16 \\
G4$-$2      &  $-$0.14   &   0.04    &   $-$0.31  &   0.09 &   $-$0.07  &   0.06  &   0.12  & $-$0.15  &   $-$0.24  &   ...    &  ...   &  0.49 \\
HIP53070    &   ...      &   ...     &   ...      &   ...   &   $-$0.06  &   ...    &   0.12  &    0.19  &   0.03     &  ...     &  ...   &  ...   \\
G204$-$30   &  $-$0.17   &   ...     &   $-$0.27  &   0.08 &    0.05    & $-$0.08 &   0.14  &    0.05  &   $-$0.09  &   ...    &  ...   &  0.37 \\
G139$-$49   &  $-$0.14   &   ...     &   $-$0.37  &   0.05 &   $-$0.14  &   0.05  &   0.36  & $-$0.15  &   $-$0.15  &   ...    &  ...   &  ...   \\
G241$-$7    &  $-$0.05   &   0.06    &   $-$0.48  &   0.07 &    0.01    &   0.01  &   0.43  &    0.05  &   $-$0.20  &   0.02  &  ...   &  ...   \\
HIP40613    &  $-$0.18   &   0.00    &   $-$0.32  &   0.06 &   $-$0.04  & $-$0.01 &   0.28  & $-$0.07  &   $-$0.15  & $-$0.26 &  0.03 &  0.24 \\
G42$-$34    &  $-$0.11   &    ...    &   $-$0.31  &   0.21 &   $-$0.10  &   0.16  &   0.02  & $-$0.14  &   $-$0.23  &   0.33  &  0.05 &  0.47 \\
HIP36491    &  $-$0.14   & $-$0.06   &   $-$0.37  &   0.24 &    0.02    & $-$0.12 &   0.20  & $-$0.03  &   $-$0.17  &   0.15  &  0.46 &  0.34 \\
HIP94931    &  $-$0.14   &   0.04    &   $-$0.20  &   0.06 &   $-$0.05  &   0.17  &   0.16  &    0.05  &   $-$0.21  &   0.17  &  0.16 &  0.43 \\
HIP74033    &  $-$0.12   &   0.01    &   $-$0.30  &   0.10 &   $-$0.03  & $-$0.02 &   0.32  &    0.11  &   $-$0.08  & $-$0.04 &  0.16 &  0.33 \\
G5$-$1      &  $-$0.18   & $-$0.05   &   $-$0.43  &   0.07 &   $-$0.07  & $-$0.40 &   0.08  & $-$0.08  &   $-$0.22  &   0.13  &  ...   &  0.29 \\
G102$-$44   &  $-$0.10   &   0.02    &   $-$0.35  &   0.04 &   $-$0.07  & $-$0.05 &   0.20  & $-$0.09  &   $-$0.14  &   0.11  &  0.40 & $-$0.01 \\ 
HIP58253    &  $-$0.14   & $-$0.01   &   $-$0.31  &   0.23 &   $-$0.00  &   0.01  &   0.13  & $-$0.02  &   $-$0.09  &   ...    &  ...   &  0.13 \\
            &            &           &             &         &             &          &          &           &             &          &        &        \\

\multicolumn{13}{c}{AF06 stream}\\

            &            &           &             &         &             &          &          &           &             &          &        &        \\ 
G67$-$40    &  $-$0.11  & $-$0.06  &   $-$0.28  &   0.08  &    0.04  &   0.14  &   0.15  & $-$0.02  &   $-$0.23  &   0.17  &  ...   &  0.36 \\
HIP9080     &  $-$0.07  & $-$0.06  &   $-$0.18  &   0.03  & $-$0.02  &   0.23  &   0.08  &    0.07  &   $-$0.10  &   0.29  &  0.22  & $-$0.01 \\
G66$-$51    &  $-$0.02  &    ...   &   $-$0.26  &   0.12  &  $-$0.02 &   0.05  &   0.05  &    0.08  &   $-$0.17  &   0.25  &  ...   &  ...   \\
G106$-$8    &  $-$0.11  & $-$0.04  &   $-$0.25  &   0.13  &    0.00  &   0.06  &   0.20  & $-$0.07  &   $-$0.08  &   0.14  &  0.12  &  0.18 \\
HIP10652    &  $-$0.07  & $-$0.01  &   $-$0.30  &   0.08  &    0.07  &   0.05  &   0.12  &    0.01  &   $-$0.20  &   0.03  &  0.22  &  0.22 \\
HIP22020    &  $-$0.03  &   0.02   &   $-$0.23  &   0.12  &    0.03  &   0.16  &   0.21  & $-$0.05  &   $-$0.07  &   0.04  &  0.17  &  0.14 \\
HIP26452    &  $-$0.09  &   0.01   &   $-$0.29  & $-$0.04 & $-$0.02  & $-$0.19 &   0.14  & $-$0.07  &   $-$0.05  &   0.22  &  0.48  &  0.09 \\
HIP31740    &  $-$0.05  &   0.03   &   $-$0.26  &   0.04  &    0.04  &   0.21  &   0.25  &    0.07  &   $-$0.13  &   0.02  &  0.15  &  0.45 \\
HIC102923   &  $-$0.02  &   0.01   &   $-$0.08  &   0.17  &    0.02  &   0.27  &   0.22  & $-$0.11  &   $-$0.21  &   0.19  &  ...   &  0.39 \\
G146$-$76   &  $-$0.25  &   ...    &   $-$0.56  &    ...  & $-$0.06  &    ...  &   0.02  & $-$0.23  &    0.05    &   ...   &  ...   &  0.23 \\
HIP34642    &  $-$0.10  & $-$0.06  &   $-$0.19  &   0.16  &    0.05  &   0.07  &   0.16  & $-$0.04  &    0.00    & $-$0.09 &  0.28  &  0.22 \\
G10$-$12    &  $-$0.05  &   0.10   &   $-$0.26  &   0.15  &    0.05  &   0.17  &   0.25  &    0.06  &   $-0$.12  & $-$0.01 &  0.24  &  0.41 \\
HIP17147    &  $-$0.12  &   0.02   &   $-$0.41  &   0.08  &    0.01  & $-$0.18 &   0.27  &    0.14  &    0.09    &   0.03  &  0.10  &  0.30 \\
HIP24030    &  $-$0.03  &    ...   &   $-$0.42  &   0.12  & $-$0.01  & $-$0.26 &   0.28  &    0.01  &   $-$0.03  &   ...   &  ...   & $-$0.02 \\
HIP11952    &  ...       &  ...    &   ...      &   ...   &    ...   &   ...   &   0.21  &    0.10  &   $-$0.15  &   ...   &  ...   &  ...   \\
HIP29814    &  $-$0.06  & $-$0.20  &   $-$0.18  &   0.08  &    0.01  &   0.15  &   0.19  & $-$0.05  &   $-$0.14  &   0.18  &  0.22  &  ...   \\
G197$-$45   &   0.03    &   0.02   &   $-$0.26  & $-$0.03 &    0.02  &   0.15  &   0.09  & $-$0.23  &   $-$0.28  &   0.14  &  ...   &  0.10 \\
HIP104913   &  $-$0.14  & $-$0.13  &   $-$0.17  &   0.00  & $-$0.06  &   0.13  &   0.10  & $-$0.03  &   $-$0.11  &   0.16  &  0.27  &  0.18 \\
G192$-$21   &  $-$0.08  & $-$0.17  &   $-$0.38  &   0.12  &    0.05  &   0.10  &   0.15  & $-$0.03  &   $-$0.08  &   ...   &  ...   &  0.12 \\
G69$-$21    &  $-$0.08  &   0.01   &   $-$0.16  &   0.05  & $-$0.01  &   0.13  &   0.09  & $-$0.02  &   $-$0.08  &   0.09  &  0.29  &  0.15 \\
G68$-$10    &  $-$0.08  & $-$0.06  &   $-$0.31  &   0.11  &    0.02  &   0.05  &   0.29  & $-$0.05  &   $-$0.16  & $-$0.16 & $-$0.07 & $-$0.03 \\
G30$-$46    &  $-$0.10  & $-$0.03  &   $-$0.13  & $-$0.02 & $-$0.10  &   0.15  & $-$0.08 & $-$0.09  &   $-$0.26  &   0.08  &  0.12   &  0.07 \\
HIP16169    &  $-$0.09  & $-$0.02  &   $-$0.23  &   0.14  &    0.03  &   0.06  &   0.07  &    0.01  &   $-$0.19  &   0.16  &  0.04   &  0.07 \\
G6$-$16     &  $-$0.04  &   0.03   &   $-$0.14  & $-$0.10 & $-$0.05  &   0.11  & $-$0.02 & $-$0.12  &   $-$0.07  &   0.00  &  0.10   & $-$0.01 \\
G78$-$41    &  $-$0.09  & $-$0.03  &   $-$0.29  &   0.08  &    0.01  &   0.21  &   0.20  & $-$0.04  &   $-$0.18  &   0.07  &  0.18   &  0.28 \\
G25$-$5     &  $-$0.04  &   0.03   &   $-$0.19  &   0.09  &    0.03  &   0.12  &   0.12  & $-$0.04  &   $-$0.14  &   0.05  &  0.36   &  0.29 \\
\hline
\end{tabular}
\end{minipage}
\end{table*}

\begin{table*}
 \centering
 \begin{minipage}{140mm}
  \caption{Error Analysis of the star HIP40613}
  \begin{tabular}{@{}lccccccc@{}}
  \hline
 Quantity  &  N   &   $\Delta$T$_{\rm eff}$  &  $\Delta$log $g$   &  $\Delta\xi_{\rm t}$  &   $\Delta$[M/H] 
 &  $\Delta$W$_{\lambda}$  &   $\sigma_{\rm model}$   \\
           &      & $\pm$50 K  &  $\pm$0.2 dex   &   $\pm$ 0.2 km/s   &   $\pm$ 0.1 dex & $\pm$ 2m\AA   &     \\
\hline
$\Delta$ [OI/Fe]   &  3 &  $\pm$ 0.05  &   $\pm$ 0.04  &    $\pm$ 0.01   &   $\pm$ 0.01  &   $\pm$ 0.02  &    $\pm$ 0.07  \\
$\Delta$ [NaI/Fe]  &  2 &  $\pm$ 0.03  &   $\pm$ 0.01  &    $\pm$ 0.00   &   $\pm$ 0.00  &   $\pm$ 0.03  &    $\pm$ 0.04  \\
$\Delta$ [MgI/Fe]  &  2 &  $\pm$ 0.03  &   $\pm$ 0.03  &    $\pm$ 0.02   &   $\pm$ 0.00  &   $\pm$ 0.02  &    $\pm$ 0.05  \\
$\Delta$ [AlI/Fe]  &  5 &  $\pm$ 0.02  &   $\pm$ 0.03  &    $\pm$ 0.01   &   $\pm$ 0.00  &   $\pm$ 0.01  &    $\pm$ 0.04  \\
$\Delta$ [SiI/Fe]  &  7 &  $\pm$ 0.01  &   $\pm$ 0.01  &    $\pm$ 0.01   &   $\pm$ 0.00  &   $\pm$ 0.01  &    $\pm$ 0.02  \\
$\Delta$ [CaI/Fe]  &  5 &  $\pm$ 0.04  &   $\pm$ 0.03  &    $\pm$ 0.03   &   $\pm$ 0.00  &   $\pm$ 0.01  &    $\pm$ 0.06  \\
$\Delta$ [ScII/Fe] &  3 &  $\pm$ 0.01  &   $\pm$ 0.08  &    $\pm$ 0.02   &   $\pm$ 0.02  &   $\pm$ 0.03  &    $\pm$ 0.09  \\
$\Delta$ [TiI/Fe]  &  7 &  $\pm$ 0.05  &   $\pm$ 0.02  &    $\pm$ 0.02   &   $\pm$ 0.00  &   $\pm$ 0.01  &    $\pm$ 0.06  \\
$\Delta$ [VI/Fe]   &  5 &  $\pm$ 0.06  &   $\pm$ 0.01  &    $\pm$ 0.00   &   $\pm$ 0.00  &   $\pm$ 0.03  &    $\pm$ 0.07  \\
$\Delta$ [CrI/Fe]  &  4 &  $\pm$ 0.05  &   $\pm$ 0.01  &    $\pm$ 0.01   &   $\pm$ 0.00  &   $\pm$ 0.02  &    $\pm$ 0.06  \\
$\Delta$ [CrII/Fe] &  1 &  $\pm$ 0.02  &   $\pm$ 0.08  &    $\pm$ 0.02   &   $\pm$ 0.02  &   $\pm$ 0.08  &    $\pm$ 0.12  \\
$\Delta$ [MnI/Fe]  &  3 &  $\pm$ 0.06  &   $\pm$ 0.01  &    $\pm$ 0.01   &   $\pm$ 0.00  &   $\pm$ 0.01  &    $\pm$ 0.06  \\
$\Delta$ [FeI/H]   & 45 &  $\pm$ 0.04  &   $\pm$ 0.01  &    $\pm$ 0.03   &   $\pm$ 0.01  &   $\pm$ 0.00  &    $\pm$ 0.05  \\
$\Delta$ [FeII/H]  &  6 &  $\pm$ 0.02  &   $\pm$ 0.08  &    $\pm$ 0.04   &   $\pm$ 0.02  &   $\pm$ 0.01  &    $\pm$ 0.09  \\
$\Delta$ [CoI/Fe]  &  3 &  $\pm$ 0.04  &   $\pm$ 0.00  &    $\pm$ 0.01   &   $\pm$ 0.00  &   $\pm$ 0.02  &    $\pm$ 0.05  \\
$\Delta$ [NiI/Fe]  & 16 &  $\pm$ 0.03  &   $\pm$ 0.01  &    $\pm$ 0.02   &   $\pm$ 0.00  &   $\pm$ 0.01  &    $\pm$ 0.04  \\
$\Delta$ [CuI/Fe]  &  3 &  $\pm$ 0.04  &   $\pm$ 0.02  &    $\pm$ 0.04   &   $\pm$ 0.01  &   $\pm$ 0.04  &    $\pm$ 0.07  \\
$\Delta$ [ZnI/Fe]  &  2 &  $\pm$ 0.01  &   $\pm$ 0.02  &    $\pm$ 0.05   &   $\pm$ 0.02  &   $\pm$ 0.03  &    $\pm$ 0.07  \\
$\Delta$ [YII/Fe]  &  4 &  $\pm$ 0.01  &   $\pm$ 0.08  &    $\pm$ 0.04   &   $\pm$ 0.02  &   $\pm$ 0.02  &    $\pm$ 0.09  \\
$\Delta$ [BaII/Fe] &  3 &  $\pm$ 0.02  &   $\pm$ 0.03  &    $\pm$ 0.12   &   $\pm$ 0.03  &   $\pm$ 0.01  &    $\pm$ 0.13  \\
$\Delta$ [CeII/Fe] &  1 &  $\pm$ 0.01  &   $\pm$ 0.08  &    $\pm$ 0.01   &   $\pm$ 0.03  &   $\pm$ 0.13  &    $\pm$ 0.16  \\
$\Delta$ [NdII/Fe] &  1 &  $\pm$ 0.02  &   $\pm$ 0.09  &    $\pm$ 0.01   &   $\pm$ 0.03  &   $\pm$ 0.16  &    $\pm$ 0.19  \\
$\Delta$ [EuII/Fe] &  2 &  $\pm$ 0.01  &   $\pm$ 0.09  &    $\pm$ 0.01   &   $\pm$ 0.03  &   $\pm$ 0.07  &    $\pm$ 0.12  \\
\hline
\end{tabular}
\end{minipage}
\end{table*}

\begin{table*}
 \centering
 \begin{minipage}{140mm}
  \caption{Comparison of current study with the Reddy et al. (2006) sample for common stars}
  \begin{tabular}{@{}lrrrrrrrr @{}}
  \hline
 Quantity  &   HIP40613   &   HIP74033  &   HIP9080  &  HIP10652  &  HIP22020  &
 HIP24030  &   HIP34642  &  Mean (dex)  \\
\hline
     $\Delta$ Teff      &    0.00   &     116.00   &       88.00   &    81.00   &     86.00 & $-$8.00  &  53.00  &      59 $\pm$ 47  \\
     $\Delta$ logg      & $-$0.14   &       0.09   &     $-$0.20   &  $-$0.16   &   $-$0.12 & $-$0.44  & $-$0.07 & $-$0.15 $\pm$ 0.16 \\   
     $\Delta$ [FeI/H]   &    0.08   &       0.15   &        0.20   &     0.08   &      0.13 & $-$0.07  &   0.02  &    0.08 $\pm$ 0.09 \\
     $\Delta$ [O/Fe]    & $-$0.08   &    $-$0.20   &     $-$0.33   &  $-$0.15   &   $-$0.15 &    0.04  & $-$0.06 & $-$0.13 $\pm$ 0.12 \\   
     $\Delta$ [Na/Fe]   & $-$0.06   &    $-$0.07   &     $-$0.02   &     0.02   &   $-$0.09 & $-$0.07  & $-$0.01 & $-$0.04 $\pm$ 0.04 \\  
     $\Delta$ [Mg/Fe]   & $-$0.02   &    $-$0.06   &     $-$0.06   &     0.05   &   $-$0.03 &    0.07  &   0.15  &    0.01 $\pm$ 0.08 \\
     $\Delta$ [Al/Fe]   & $-$0.01   &      ...     &     $-$0.09   &     0.05   &   $-$0.04 &    0.08  & $-$0.01 &    0.00 $\pm$ 0.06 \\
     $\Delta$ [Si/Fe]   & $-$0.05   &    $-$0.11   &     $-$0.12   &     0.06   &   $-$0.05 &    0.00  &   0.00  & $-$0.04 $\pm$ 0.06 \\   
     $\Delta$ [Ca/Fe]   &    0.02   &    $-$0.12   &       ...     &     0.09   &   $-$0.01 &    0.12  &   0.06  &    0.03 $\pm$ 0.09 \\
     $\Delta$ [Sc/Fe]   & $-$0.13   &    $-$0.02   &     $-$0.21   &  $-$0.15   &   $-$0.03 & $-$0.15  &   0.12  & $-$0.08 $\pm$ 0.11 \\  
     $\Delta$ [Ti/Fe]   &    0.04   &    $-$0.05   &        0.08   &     0.12   &      0.05 &    0.07  &   0.11  &    0.06 $\pm$ 0.06 \\
     $\Delta$ [V/Fe]    &    0.00   &    $-$0.14   &        0.08   &     0.07   &   $-$0.01 &   ...    &   0.11  &    0.02 $\pm$ 0.09 \\ 
     $\Delta$ [Cr/Fe]   & $-$0.15   &      ...     &     $-$0.02   &  $-$0.06   &      0.00 &    0.03  & $-$0.06 & $-$0.04 $\pm$ 0.06 \\
     $\Delta$ [Mn/Fe]   & $-$0.01   &       0.05   &     $-$0.14   &     0.03   &      0.00 &    0.12  &   0.08  &    0.02 $\pm$ 0.08 \\
     $\Delta$ [Co/Fe]   & $-$0.03   &    $-$0.10   &     $-$0.07   &     0.02   &      0.03 &    0.10  &   0.14  &    0.01 $\pm$ 0.09 \\
     $\Delta$ [Ni/Fe]   &    0.00   &    $-$0.06   &     $-$0.13   &     0.05   &      0.04 & $-$0.04  &   0.07  & $-$0.01 $\pm$ 0.07 \\
     $\Delta$ [Zn/Fe]   &    0.18   &       0.14   &        0.06   &     0.02   &      0.10 &    0.15  &   0.20  &    0.12 $\pm$ 0.07 \\
     $\Delta$ [Y/Fe]    & $-$0.02   &    $-$0.06   &        0.35   &     0.08   &      0.08 & $-$0.20  &   0.12  &    0.05 $\pm$ 0.17 \\
     $\Delta$ [Ba/Fe]   &    0.04   &       0.03   &        0.01   &  $-$0.11   &   $-$0.01 & $-$0.04  &   ...   & $-$0.01 $\pm$ 0.06 \\
     $\Delta$ [Ce/Fe]   & $-$0.11   &    $-$0.06   &        0.08   &     0.10   &      0.02 &    ...   &   0.14  &    0.03 $\pm$ 0.10 \\
     $\Delta$ [Nd/Fe]   & $-$0.04   &    $-$0.03   &       ...     &     ...    &       ... &    ...   &   0.20  &    0.04 $\pm$ 0.14 \\
     $\Delta$ [Cu/Fe]   &    0.03   &       0.07   &        0.20   &     0.07   &      0.12 &     0.07 &   0.17  &    0.10 $\pm$ 0.06 \\
     $\Delta$ [Eu/Fe]   &    0.00   &    $-$0.01   &     $-$0.20   &  $-$0.17   &   $-$0.07 &    ...   &   0.07  & $-$0.06 $\pm$ 0.10 \\
 \hline
\end{tabular}
\end{minipage}
\end{table*}

\begin{table*}
 \centering
 \begin{minipage}{180mm}
  \caption{The kinematical parameters of the sample.}
  \begin{tabular}{@{}llllrcccclcl@{}}
  \hline
Star  &   R$_{\rm V}$  &   U$_{\rm LSR}\pm1\sigma$   &   V$_{\rm LSR}\pm1\sigma$   &   W$_{\rm LSR}\pm1\sigma$  &  R$_{\rm m}$  &
 \it{e}    &   Z$_{\rm max}$  &   Age   &  P    &   J$_{\rm z}$    &    J$_{\rm \bot}$   \\
     & km s$^{-1}$ &  km s$^{-1}$ &  km s$^{-1}$ &  km s$^{-1}$  &  kpc  &     &     kpc   &  Gyr  &   \%   &  kpc km s$^{-1}$   &    kpc km s$^{-1}$  \\
\hline
           &            &                    &                      &                     &       &      &       &                         &             &          &    \\
\multicolumn{12}{c}{Arcturus stream} \\
           &            &                    &                      &                     &       &      &       &                         &             &          &    \\
 HIP105888 &  $-$84.3  & $-$23.9 $\pm$ 0.7  &  $-$122.4 $\pm$ 4.8  &  $-$40.0 $\pm$ 6.5  &  5.42 & 0.57 &  0.35 &   10.8$_{-0.8}^{+0.9}$  & 97 $\pm$ 0  &  $-$825  &  338\\
  HIP36710 &  $-$70.8  & $+$45.8 $\pm$ 1.3  &  $-$107.3 $\pm$ 6.9  &  $+$45.3 $\pm$ 7.0  &  5.82 & 0.51 &  0.42 &   ...                   & 97 $\pm$ 1  &  $-$965  &  388 \\
  HIP77637 &  $-$51.0   & $-$26.9 $\pm$ 1.2  &  $-$134.0 $\pm$ 17.7 &  $+$15.8 $\pm$ 6.1  &  5.21 & 0.63 &  0.14 &   ...                   & 96 $\pm$ 0  &  $-$724  &  131 \\
 G103$-$53 & $+$10.0   & $+$10.0 $\pm$ 1.8  &  $-$125.4 $\pm$ 19.1 &  $-$55.4 $\pm$ 9.6  &  5.44 & 0.58 &  0.54 &   ...                   & 96 $\pm$ 3  &  $-$813  &  476 \\
  G72$-$12 &  $-$34.3  & $-$13.7 $\pm$ 6.6  &  $-$118.7 $\pm$ 15.0 &  $-$62.9 $\pm$ 12.7 &  5.55 & 0.55 &  0.65 &   ...                   & 96 $\pm$ 2  &  $-$866  &  538\\
    G4$-$2 &  $+$38.8  & $-$11.1 $\pm$ 1.2  &  $-$117.2 $\pm$ 19.8 &  $-$76.0 $\pm$ 8.1  &  5.60 & 0.54 &  0.90 &   ...                   & 95 $\pm$ 4  &  $-$880  &  650\\
  HIP53070 &  $+$65.8  & $-$23.6 $\pm$ 0.71 &  $-$124.8 $\pm$ 4.84 &  $+$22.9 $\pm$ 2.1  &  5.41 & 0.59 &  0.19 &   ...                   & 95 $\pm$ 1  &  $-$811  &  194\\
 G204$-$30 &  $-$69.7  & $+$58.9 $\pm$ 11.5 &  $-$130.4 $\pm$ 12.0 &  $+$46.9 $\pm$ 10.7 &  5.43 & 0.62 &  0.44 &   ...                   & 95 $\pm$ 3  &  $-$763  &  399 \\
 G139$-$49 &  $-$94.1  & $-$25.9 $\pm$ 5.5  &  $-$131.7 $\pm$ 11.8 &   $-$6.7 $\pm$ 1.9  &  5.24 & 0.62 &  0.06 &   ...                   & 95 $\pm$ 1  &  $-$742  &   57 \\
  G241$-$7 & $-$113.6  & $-$45.5 $\pm$ 14.1 &  $-$134.1 $\pm$ 5.3  &  $-$20.4 $\pm$ 1.5  &  5.34 & 0.64 &  0.16 &   ...                   & 95 $\pm$ 0  &  $-$729  &  175 \\
  HIP40613 &  $+$112.5 & $-$31.6 $\pm$ 1.7  &  $-$139.0 $\pm$ 3.3  &  $-$32.7 $\pm$ 3.5  &  5.21 & 0.66 &  0.27 &   12.9$_{-0.3}^{+0.5}$  & 95 $\pm$ 1  &  $-$692  &  279\\
  G42$-$34 & $+$36.9   &  $+$8.4 $\pm$ 2.7  &  $-$126.5 $\pm$ 17.6 &   $-$7.1 $\pm$ 6.5  &  5.37 & 0.59 &  0.08 &   ...                   & 94 $\pm$ 6  &  $-$799  &   60 \\
  HIP36491 &   $+$90.8 & $-$47.7 $\pm$ 1.8  &  $-$118.9 $\pm$ 6.1  &  $+$1.2 $\pm$ 2.2   &  5.60 & 0.57 &  0.01 &   ...                   & 91 $\pm$ 4  &  $-$865  &   10\\
  HIP94931 &  $-$121.4 & $+$65.9 $\pm$ 2.7  &  $-$120.7 $\pm$ 1.2  &  $-$79.2 $\pm$ 1.8  &  5.70 & 0.58 &  0.95 &   13.4$_{-0.1}^{+3.2}$  & 91 $\pm$ 1  &  $-$845  &  672\\
  HIP74033 &  $-$59.8  &$-$111.7 $\pm$ 7.5  &  $-$132.7 $\pm$ 11.5 &  $+$42.3 $\pm$ 7.2  &  5.83 & 0.68 &  0.40 &   12.0$_{-1.3}^{+1.6}$  & 89 $\pm$ 5  &  $-$738  &  351 \\
    G5$-$1 &  $-$22.4  & $+$44.8 $\pm$ 2.4  &  $-$125.7 $\pm$ 15.5 &  $-$85.7 $\pm$ 13.2 &  5.56 & 0.59 &  1.07 &   ...                   & 89 $\pm$ 8  &  $-$810  &  737 \\
 G102$-$44 &  $-$29.2  & $+$73.6 $\pm$ 5.4  &  $-$130.5 $\pm$ 21.3 &  $+$88.1 $\pm$ 11.7 &  5.68 & 0.63 &  1.15 &   ...                   & 82 $\pm$ 14 &  $-$769  &  759 \\
  HIP58253 &  $+$29.3  & $-$62.5 $\pm$ 7.5  &  $-$106.2 $\pm$ 10.8 &   $-$3.4 $\pm$ 4.2  &  5.89 & 0.52 &  0.08 &   ...                   & 80 $\pm$ 16 &  $-$969  &   34 \\
           &            &                    &                      &                     &       &      &       &                         &             &          &    \\
\multicolumn{12}{c}{AF06 stream} \\
           &            &                    &                      &                     &       &      &       &                         &             &          &    \\
  G67$-$40 &  $-$29.2  & $-$94.8 $\pm$ 15.3 &   $-$84.5 $\pm$ 10.2 & $-$54.8 $\pm$ 12.2  &  6.61 & 0.48 &  0.57 &   ...                   & 97 $\pm$ 1  &  $-$1144 &   459\\
   HIP9080 &  $-$10.2  & $-$92.8 $\pm$ 20.9 &   $-$91.2 $\pm$ 17.9 & $+$54.2 $\pm$ 7.7   &  6.49 & 0.49 &  0.57 &   ...                   & 97 $\pm$ 1  &  $-$1098 &   469\\
  G66$-$51 &  $-$118.5 & $-$86.1 $\pm$ 2.5  &   $-$79.8 $\pm$ 10.5 & $-$66.9 $\pm$ 2.2   &  6.62 & 0.44 &  0.76 &   ...                   & 97 $\pm$ 0  &  $-$1183 &    571  \\
  G106$-$8 &  $+$49.4  & $-$16.3 $\pm$ 2.5  &   $-$92.2 $\pm$ 11.0 & $+$85.9 $\pm$ 13.4  &  6.12 & 0.42 &  1.05 &   14.0$_{-1.3}^{+1}$    & 96 $\pm$ 2  &  $-$1097 &   737\\
  HIP10652 &  $-$20.9  & $-$78.2 $\pm$ 7.7  &   $-$79.1 $\pm$ 5.7  & $+$84.7 $\pm$ 4.9   &  6.68 & 0.42 &  1.08 &   ...                   & 96 $\pm$ 1  &  $-$1201 &   726\\
  HIP22020 &  $+$30.6  & $-$74.3 $\pm$ 7.9  &   $-$85.6 $\pm$ 14.5 & $-$52.9 $\pm$ 8.7   &  6.46 & 0.45 &  0.53 &   11.8$_{-2}^{+1.2}$    & 96 $\pm$ 3  &  $-$1151 &   455\\
  HIP26452 &  $-$35.8  & $+$ 1.2 $\pm$ 4.6  &   $-$98.0 $\pm$ 10.3 & $+$50.0 $\pm$ 6.7   &  5.91 & 0.45 &  0.47 &   ...                   & 96 $\pm$ 3  &  $-$1045 &   428\\
  HIP31740 &  $+$86.2  & $-$94.8 $\pm$ 4.3  &  $-$107.5 $\pm$ 21.5 & $+$49.6 $\pm$ 2.7   &  6.20 & 0.56 &  0.50 &   ...                   & 96 $\pm$ 2  &  $-$964  &   423 \\
 HIP102923 &  $-$61.7  & $-$12.2 $\pm$ 1.0  &   $-$92.0 $\pm$ 3.4  & $-$52.6 $\pm$ 5.3   &  5.98 & 0.42 &  0.50 &   ...                   & 96 $\pm$ 2  &  $-$1084 &   446\\
 G146$-$76 &  $-$113.7 & $+$50.3 $\pm$ 1.5  &   $-$80.6 $\pm$ 8.6  & $-$90.6 $\pm$ 1.0   &  6.46 & 0.39 &  1.16 &   ...                   & 96 $\pm$ 1  &  $-$1189 &    770 \\
  HIP34642 &  $-$28.1  & $-$13.1 $\pm$ 5.4  &   $-$85.5 $\pm$ 9.4  & $-$53.9 $\pm$ 5.6   &  6.19 & 0.39 &  0.52 &   10.6$_{-1.6}^{+1.3}$  & 95 $\pm$ 4  &  $-$1154 &   463 \\  
 G10$-$12  &  $+$133.0 & $-$52.1 $\pm$ 6.5  &   $-$91.4 $\pm$ 4.8  & $+$96.8 $\pm$ 3.8   &  6.24 & 0.43 &  1.29 &   ...                   & 94 $\pm$ 1  &  $-$1095 &    822 \\ 
  HIP17147 & $+$120.0  & $-$101.3 $\pm$ 0.9 &   $-$83.8 $\pm$ 1.2  & $-$34.2 $\pm$ 1.0   &  6.70 & 0.48 &  0.31 &   ...                   & 88 $\pm$ 1  &  $-$1160 &   290\\
  HIP24030 &  $-$15.9  & $+$21.4 $\pm$ 1.2  &  $-$105.6 $\pm$ 23.5 & $+$112.3 $\pm$ 20.5 &  5.94 & 0.47 &  1.65 &   ...                   & 85 $\pm$ 16 &  $-$984  &   966\\
 HIP11952  &  $+$24.0  & $+$38.0 $\pm$ 6.47 &   $-$90.7 $\pm$ 15.3 & $-$30.7 $\pm$ 2.9   &  6.11 & 0.43 &  0.29 &   ...                   & 72 $\pm$ 27 &  $-$1106 &    267 \\
  HIP29814 &  $+$22.1  & $-$43.2 $\pm$ 2.1  &   $-$92.8 $\pm$ 6.0  & $-$20.9 $\pm$ 2.0   &  6.08 & 0.45 &  0.17 &   ...                   & 62 $\pm$ 13 &  $-$1087 &   179\\
 G197$-$45 &  $+$22.8  & $-$67.0 $\pm$ 10.4 &   $-$76.9 $\pm$ 13.2 & $+$32.1 $\pm$ 1.2   &  6.55 & 0.40 &  0.29 &   ...                   & 61 $\pm$ 24 &  $-$1219 &   268 \\
 HIP104913 &  $-$64.5  & $-$69.6 $\pm$ 5.4  &   $-$80.4 $\pm$ 1.8  & $+$26.9 $\pm$ 1.9   &  6.47 & 0.42 &  0.23 &   11.9$_{-2.6}^{+1.9}$  & 58 $\pm$ 6  &  $-$1184 &   228\\
 G192$-$21 &  $-$18.6  & $+$ 7.0 $\pm$ 3.2  &   $-$88.4 $\pm$ 13.1 & $+$22.1 $\pm$ 3.0   &  6.08 & 0.41 &  0.18 &   11.2$_{-2.9}^{+2.1}$  & 47 $\pm$ 28 &  $-$1125 &   189\\
  G69$-$21 &  $-$15.8  & $-$87.8 $\pm$ 15.6 &   $-$80.9 $\pm$ 11.1 & $+$ 3.2 $\pm$ 2.0   &  6.63 & 0.45 &  0.06 &   ...                   & 45 $\pm$ 24 &  $-$1182 &    33\\
  G68$-$10 &  $-$40.6  & $-$86.1 $\pm$ 8.6  &   $-$76.8 $\pm$ 4.6  & $-$16.0 $\pm$ 4.4   &  6.67 & 0.43 &  0.14 &   13.2$_{-1.5}^{+1.7}$  & 45 $\pm$ 11 &  $-$1213 &   133\\
  G30$-$46 &  $-$22.9  & $-$75.1 $\pm$ 13.4 &   $-$80.8 $\pm$ 10.6 & $-$12.9 $\pm$ 5.5   &  6.50 & 0.43 &  0.13 &   ...                   & 42 $\pm$ 22 &  $-$1181 &   105\\
  HIP16169 &  $+$63.3  & $-$47.4 $\pm$ 1.0  &   $-$87.5 $\pm$ 3.6  & $-$10.5 $\pm$ 1.4   &  6.20 & 0.42 &  0.09 &  12.85$_{-0.1}^{+1.05}$ & 40 $\pm$ 7  &  $-$1131 &    88\\
   G6$-$16 &  $+$25.6  & $-$31.4 $\pm$ 3.0  &   $-$85.4 $\pm$ 14.3 & $-$15.9 $\pm$ 1.9   &  6.21 & 0.40 &  0.13 &   11.0$_{-1.5}^{+1.1}$  & 36 $\pm$ 27 &  $-$1154 &   135\\
  G78$-$41 &  $-$10.3  & $-$18.7 $\pm$ 5.7  &   $-$88.7 $\pm$ 13.3 &  $-$3.3 $\pm$ 2.3   &  6.11 & 0.41 &  0.03 &   ...                   & 32 $\pm$ 25 &  $-$1127 &    28\\
   G25$-$5 &  $-$38.1  & $+$44.2 $\pm$ 8.4  &   $-$85.0 $\pm$ 9.5  &  $-$2.4 $\pm$ 3.9   &  6.17 & 0.41 &  0.04 &   ...                   & 31 $\pm$ 17 &  $-$1143 &    23\\
 \hline
\end{tabular}
\end{minipage}
\end{table*}

\begin{table*}
 \centering
 \begin{minipage}{140mm}
  \caption{Mean elemental abundance ratios and dispersions}
  \begin{tabular}{@{}lccccccc@{}}
  \hline
\multicolumn{1}{c}{Element}& \multicolumn{2}{c}{Arcturus} & \multicolumn{2}{c}{AF06} & \multicolumn{2}{c}{Thick disk} & 
\multicolumn{1}{c}{$\sigma_{\rm model}$}\\
       &  Mean  &  $\sigma$  &  Mean  &  $\sigma$   &   Mean  & $\sigma$  &   \\
\hline
{}[O/Fe]    &   0.54   &    0.13  &   0.48   &   0.08 &   0.60   &        0.12  & 0.07 \\   
{}[Na/Fe]   &   0.03   &    0.09  &   0.07   &   0.04 &   0.10   &        0.06  & 0.04 \\   
{}[Mg/Fe]   &   0.21   &    0.11  &   0.21   &   0.09 &   0.30   &        0.07  & 0.05 \\   
{}[Al/Fe]   &   0.18   &    0.04  &   0.21   &   0.03 &   0.26   &        0.08  & 0.04 \\   
{}[Si/Fe]   &   0.14   &    0.06  &   0.15   &   0.04 &   0.22   &        0.06  & 0.02 \\   
{}[Ca/Fe]   &   0.14   &    0.05  &   0.16   &   0.04 &   0.16   &        0.06  & 0.06 \\   
{}[Sc/Fe]   &   0.07   &    0.09  &   0.07   &   0.07 &   0.14   &        0.09  & 0.09 \\   
{}[TiI/Fe]  &   0.20   &    0.05  &   0.23   &   0.04 &   0.19   &        0.07  & 0.06 \\   
{}[V/Fe]    &   0.07   &    0.08  &   0.11   &   0.06 &   0.11   &        0.08  & 0.07 \\   
{}[CrI/Fe]  &  -0.13   &    0.03  &  -0.07   &   0.04 &  -0.03   &        0.04  & 0.06 \\   
{}[CrII/Fe] &  -0.01   &    0.05  &  -0.03   &   0.08 &   ...    &        ...   & 0.12 \\
{}[Mn/Fe]   &  -0.32   &    0.05  &  -0.27   &   0.05 &  -0.26   &        0.07  & 0.06 \\  
{}[Co/Fe]   &   0.10   &    0.07  &   0.09   &   0.05 &   0.10   &        0.05  & 0.05 \\   
{}[Ni/Fe]   &  -0.04   &    0.05  &   0.02   &   0.02 &   0.02   &        0.04  & 0.04 \\   
{}[Cu/Fe]   &   0.01   &    0.08  &   0.08   &   0.08 &  -0.02   &        0.07  & 0.07 \\   
{}[Zn/Fe]   &   0.22   &    0.09  &   0.17   &   0.07 &   0.12   &        0.06  & 0.07 \\                                       
{}[Y/Fe]    &  -0.03   &    0.08  &  -0.02   &   0.08 &   0.00   &        0.11  & 0.09 \\   
{}[Ba/Fe]   &  -0.15   &    0.06  &  -0.13   &   0.08 &  -0.14   &        0.09  & 0.13 \\   
{}[Ce/Fe]   &   0.06   &    0.15  &   0.08   &   0.11 &   0.06   &        0.13  & 0.16 \\     
{}[Nd/Fe]   &   0.22   &    0.14  &   0.19   &   0.14 &   0.20   &        0.16  & 0.19 \\   
{}[Eu/Fe]   &   0.29   &    0.15  &   0.22   &   0.13 &   0.36   &        0.12  & 0.12 \\	   
\hline
\end{tabular}
\end{minipage}
\end{table*}
 
\clearpage

\begin{figure*}
\includegraphics[width=84mm]{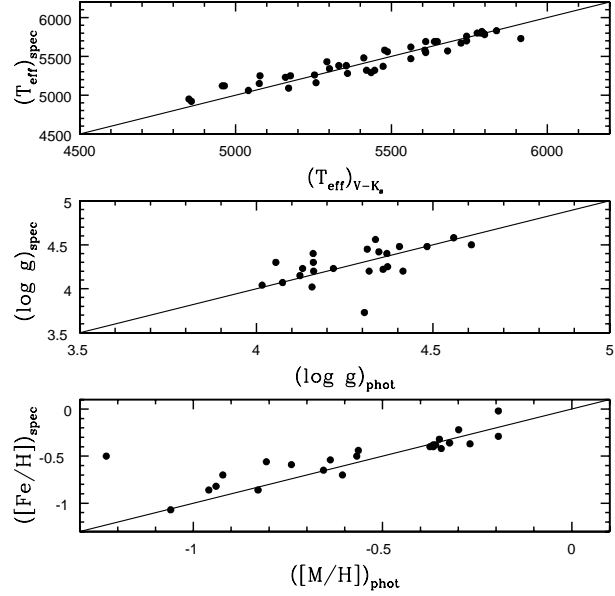}
\caption{Comparison of atmospheric parameters derived using
spectroscopy and photometry.}
\label{}
\end{figure*}

\begin{figure*}
\vspace{.5cm}
\includegraphics[width=100mm]{err.eps}
\caption Estimation of uncertainties in the T$_{\rm eff}$ (top) and $\xi_{t}$ (bottom). 
The solid center line is for the best representative model parameter. Broken lines represent
models with increaseing or decreasing T$_{\rm eff}$ and $\xi_{t}$ in steps of 25~K and 0.1 km s$^{-1}$,
respectively.  
\label{}
\end{figure*}

\begin{figure*}
\includegraphics[width=84mm]{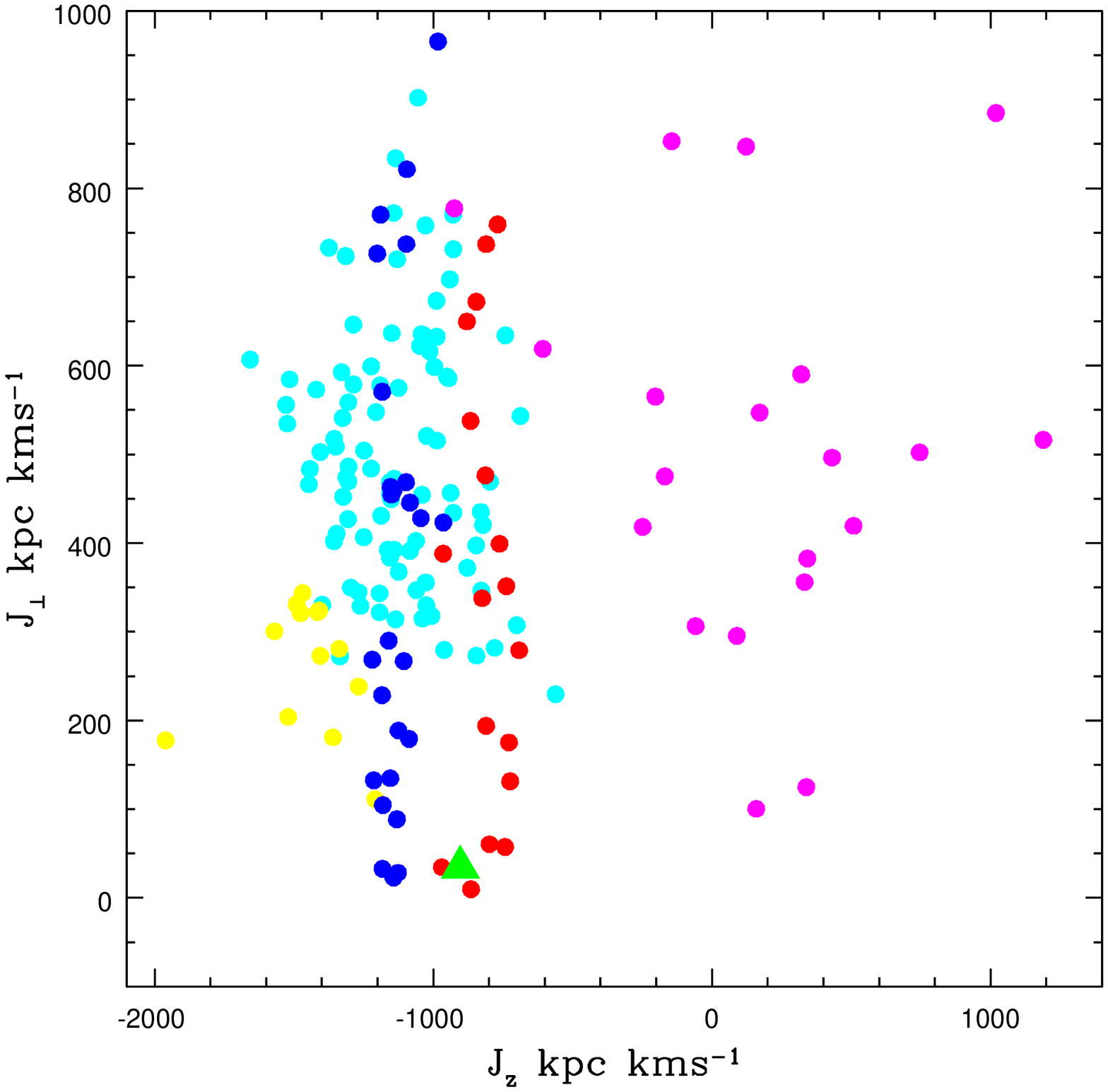}
\caption{Stream stars are shown with members of halo, thick disk and thin disk
stars taken from Reddy et al. (2006) in the angular momentum space: J$_{\bot}$ versus J$_{\rm z}$. Red :
Arcturus stream, Blue : AF06 stream, Cyan : Thick
disk, Yellow : Thin disk, Magenta : Halo, Green : Giant Arcturus. [Colour Online]}
\label{}
\end{figure*}

\begin{figure*}
\includegraphics[width=84mm]{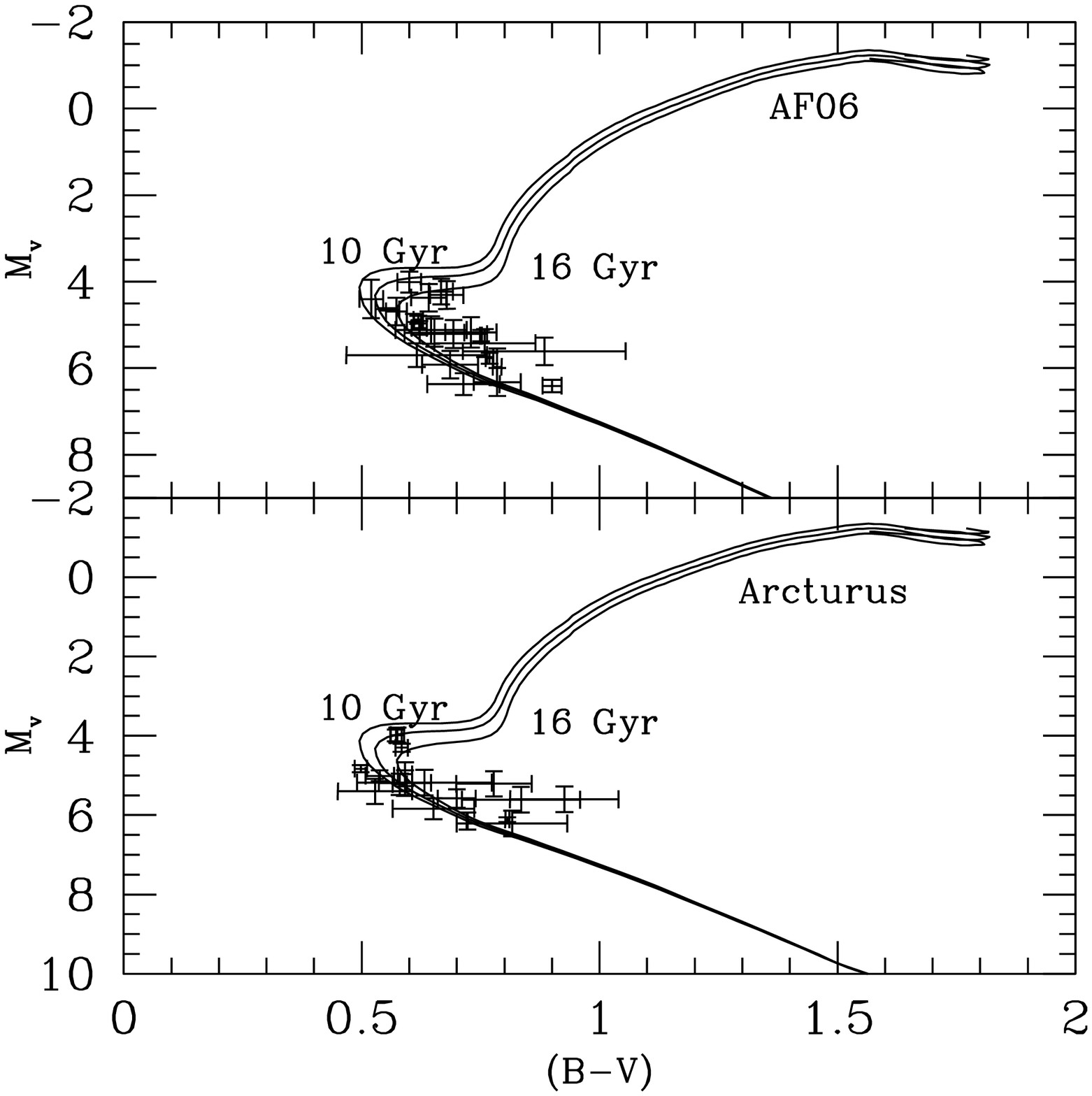}
\caption{The colour-magnitude diagram for the Arcturus and AF06 stream.}
\label{}
\end{figure*}

\begin{figure*}
\vspace{1.4cm}
\includegraphics[width=190mm]{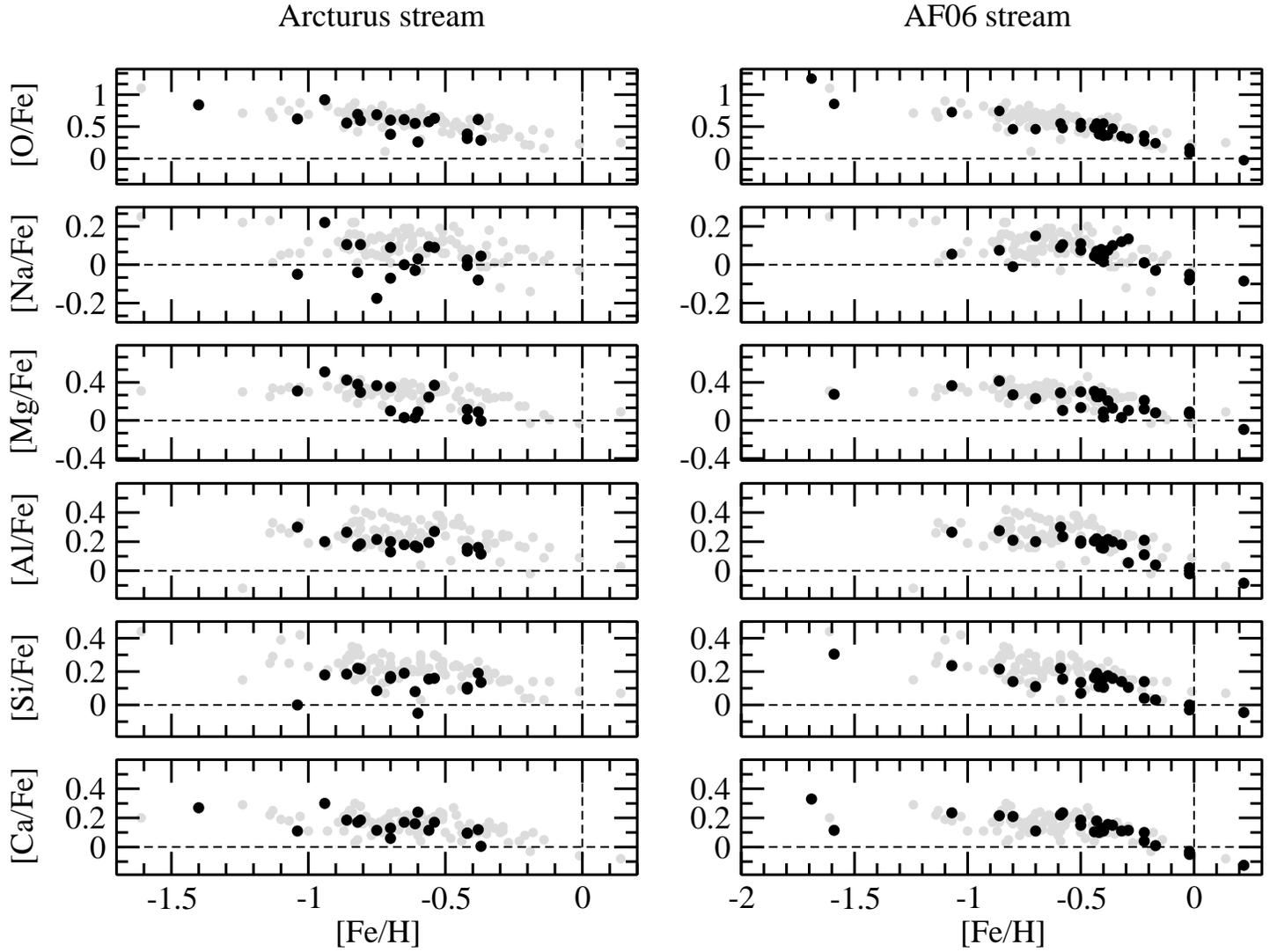}
\caption{The abundance plots of elements - O, Na, Mg, Al, Si and Ca for the 
Arcturus stream (left column) and AF06 stream (right column).
 Grey symbols represent the field thick disk members from Reddy et al. (2006). [Colour Online]}
\label{ }
\end{figure*}

\begin{figure*}
\vspace{1.4cm}
\includegraphics[width=190mm]{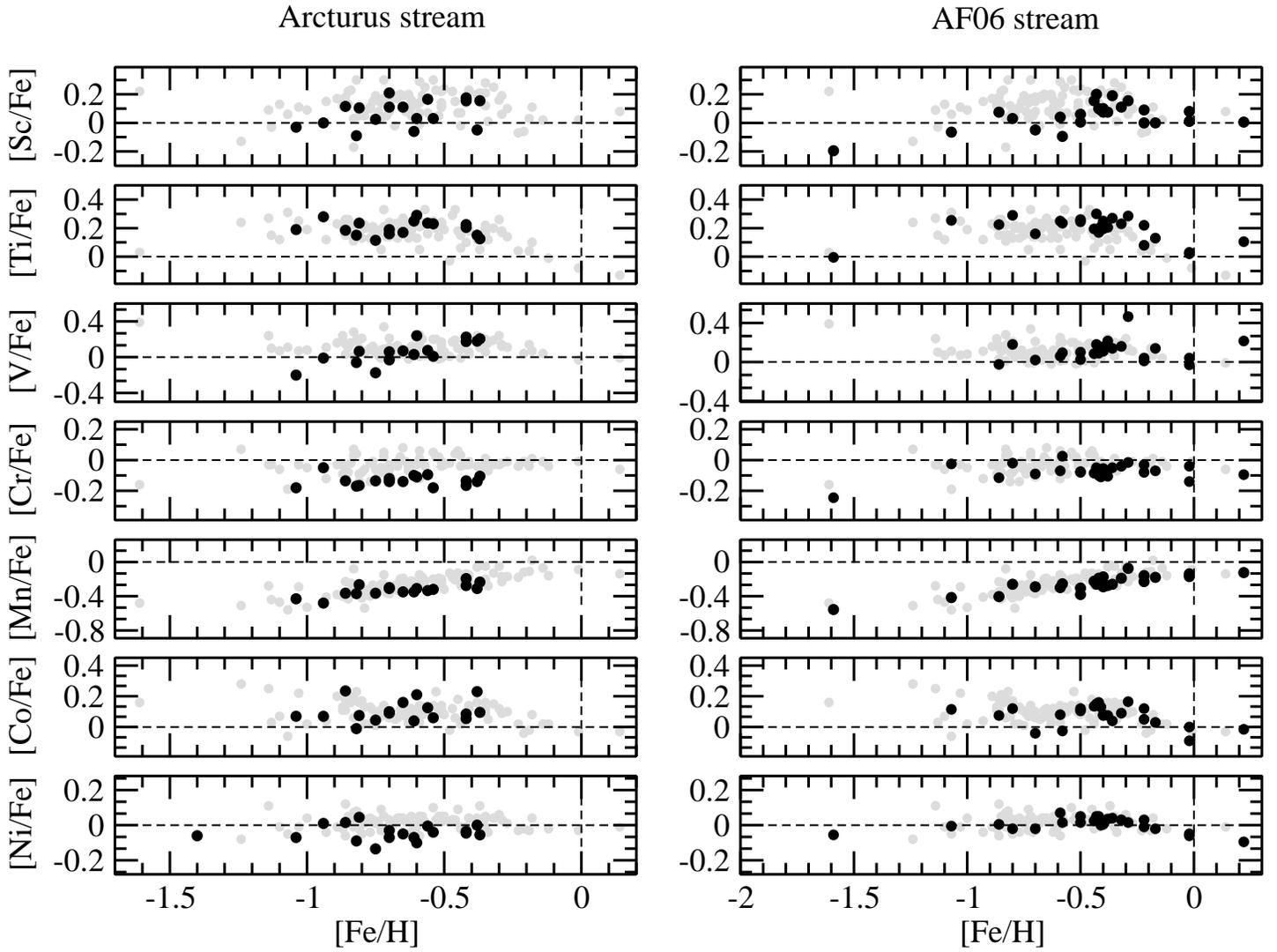}
\caption{Same as Figure~4 but for the elements - Sc, Ti, V, Cr, Mn, Co and Ni. [Colour Online]}
\label{}
\end{figure*}

\begin{figure*}
\vspace{1.5cm}
\includegraphics[width=190mm]{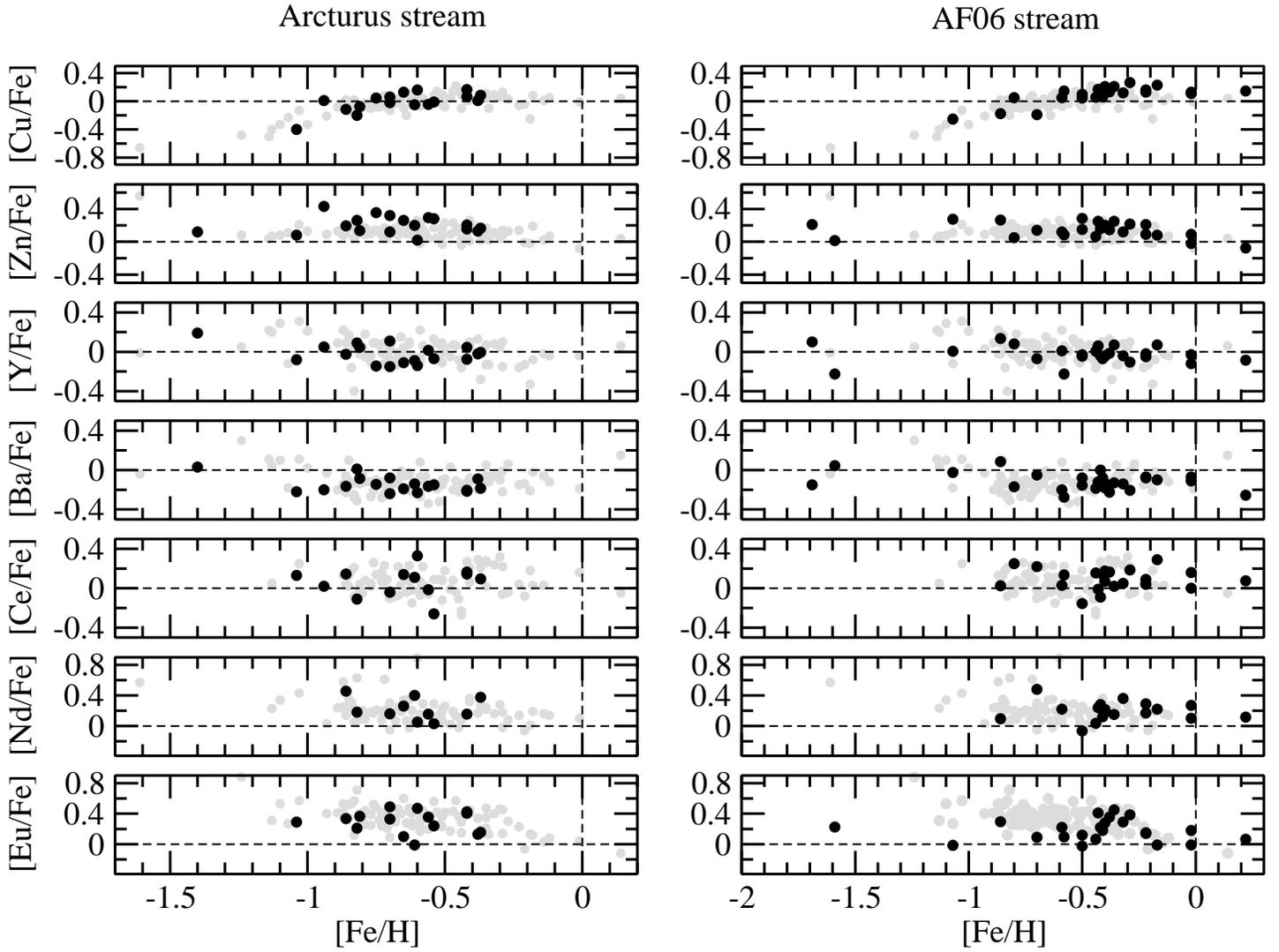}
\caption{Same as Figure~4 but for the elements - Cu, Zn, Y, Ba, Ce, Nd and Eu. [Colour Online]}
\label{}
\end{figure*}

\begin{figure*}
\includegraphics[width=84mm]{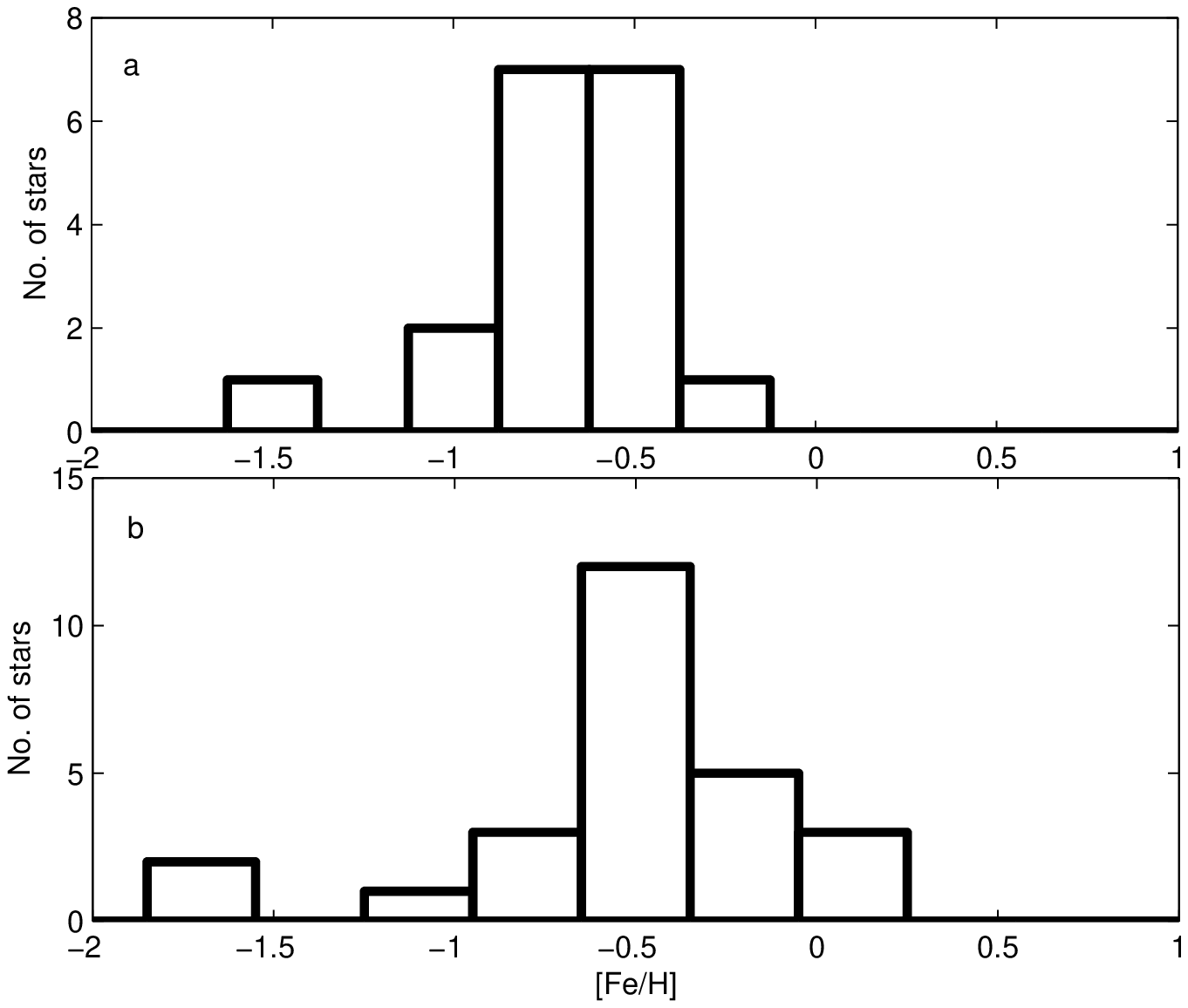}
\caption{The metallicity distribution of a) the Arcturus stream, binsize = 0.25 dex and
 b) the AF06 stream, binsize = 0.3 dex.}
\label{}
\end{figure*}

\begin{figure*}
\includegraphics[width=84mm]{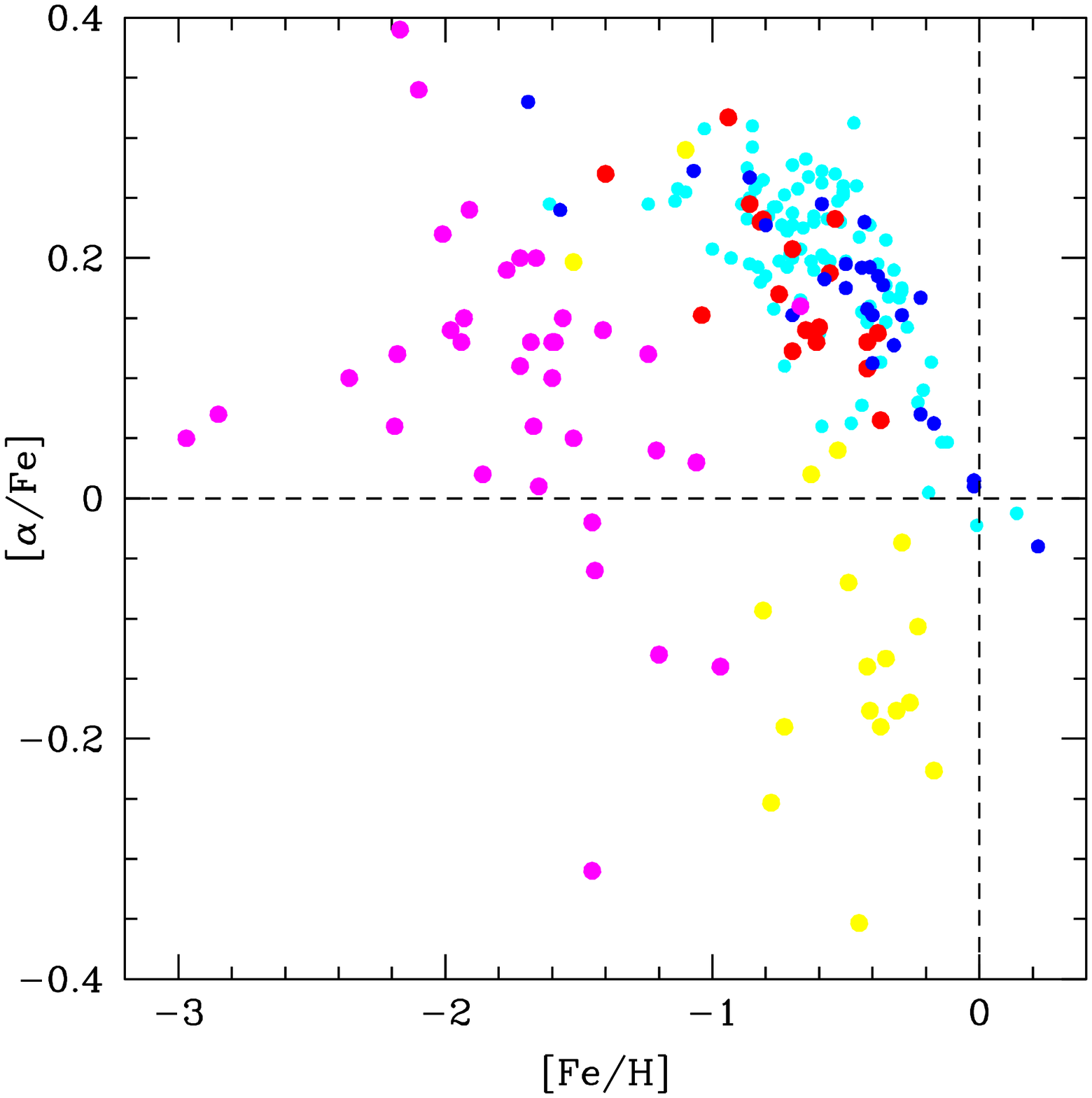}
\caption{The [$\alpha$/Fe] versus [Fe/H] plot. Red : Arcturus stream, Blue : AF06 stream, Cyan : Thick disk, Magenta : dSph satellite 
galaxies (Draco, Sculptor, Sextans, Ursa Minor, Carina, Fornax, Leo I) from Venn et al (2004), yellow : Sgr dSph from 
Monaco et al. (2005) with 
[$\alpha$/Fe] = ([Mg/Fe]+[Ca/Fe]+[Ti/Fe])/3. [Colour Online]}
\label{}
\end{figure*}

\begin{figure*}
\includegraphics[width=84mm]{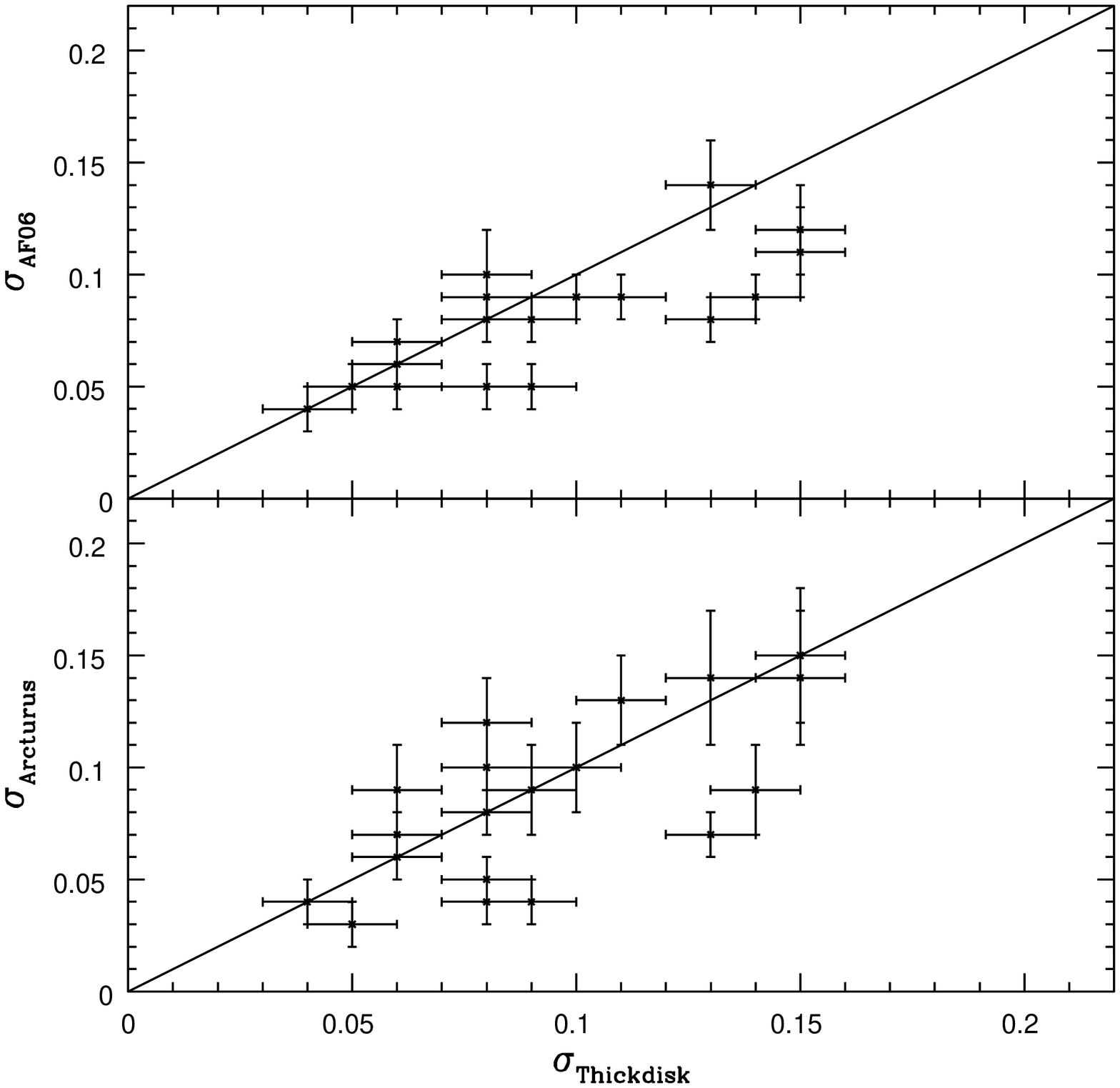}
\caption{The dispersions in [X/Fe] of members of the Arcturus stream, 
the AF06 stream and the field thick disk sample}
\label{}
\end{figure*}

\label{lastpage}

\end{document}